\documentclass[pdflatex,sn-nature]{sn-jnl}

\usepackage{graphicx}%
\usepackage{multirow}%
\usepackage{amsmath,amssymb,amsfonts}%
\usepackage{amsthm}%
\usepackage{mathrsfs}%
\usepackage[title]{appendix}%
\usepackage{xcolor}%
\usepackage{textcomp}%
\usepackage{manyfoot}%
\usepackage{booktabs}%
\usepackage{algorithm}%
\usepackage{algorithmicx}%
\usepackage{algpseudocode}%
\usepackage{listings}%

\begin{document}

\title{Configurational Entropy and Adam-Gibbs Relation for Quantum Liquids}

\author[1,2]{\fnm{Yang} \sur{Zhou}}\email{yzhou4@gradcenter.cuny.edu}
\author[1,2]{\fnm{Ali} \sur{Eltareb}}\email{aeltareb@gradcenter.cuny.edu}
\author[3,4]{\fnm{Gustavo} \sur{E. Lopez}}\email{gustavo.lopez1@lehman.cuny.edu}
\author[1,2,3]{\fnm{Nicolas} \sur{Giovambattista}}\email{ngiovambattista@brooklyn.cuny.edu}

\affil[1]{\orgdiv{Ph.D. Program in Physics}, \orgname{The Graduate Center of the City University of New York}, \orgaddress{ \city{New York}, \postcode{10016}, \state{NY}, \country{United States}}}
\affil[2]{\orgdiv{Department of Physics}, \orgname{Brooklyn College of the City University of New York}, \orgaddress{ \city{New York}, \postcode{11210}, \state{NY}, \country{United States}}}
\affil[3]{\orgdiv{Ph.D. Program in Chemistry}, \orgname{The Graduate Center of the City University of New York}, \orgaddress{ \city{New York}, \postcode{10016}, \state{NY}, \country{United States}}}
\affil[4]{\orgdiv{Department of Chemistry}, \orgname{Lehman College of the City University of New York}, \orgaddress{ \city{New York}, \postcode{10468}, \state{NY}, \country{United States}}}

\date{\today}

\abstract{As a liquid approaches the glass state, its dynamics slows down rapidly, by a few orders of magnitude in a very small temperature range. In the case of light elements and small molecules containing hydrogen (e.g., water), such a process can be affected by nuclear quantum effects (due to quantum fluctuations/atoms delocalization).   In this work, we apply  the potential
energy landscape (PEL) formalism and path-integral computer simulations to study the low-temperature behavior of a Lennard-Jones binary mixture (LJBM) that obeys \textit{quantum} mechanics.    We show that, as for the case of classical liquids, (i) a configurational entropy $S_{IS}$ can be defined, and (ii) the Adam-Gibbs equation, which relates the diffusion coefficient of a liquid and its $S_{IS}$, holds for the studied \textit{quantum} LJBM. Overall, our work shows that one theoretical approach, the PEL formalism, can be used to describe low-temperature liquids close to their glass transition, independently of whether the system obeys classical or quantum mechanics.}


\maketitle



\section{Introduction}

Glasses, or amorphous solids, are out-of-equilibrium systems that can be formed by different routes, 
including cooling a liquid sufficiently fast so crystallization can be avoided ~\cite{angellFormationGlassesLiquids1995,debenedettiSupercooledLiquidsGlass2001,binderGlassyMaterialsDisordered2011,berthierFacetsGlassPhysics2016}.  The behavior of liquids close to 
the glass state, and during the associated liquid-to-glass transition have been the focus of numerous studies over
 the last few decades ~\cite{kauzmannNatureGlassyState1948,adamTemperatureDependenceCooperative1965,edigerSpatiallyHeterogeneousDynamics2000,schoenholzStructuralApproachRelaxation2016},  and various theoretical approaches have been proposed to describe these systems ~\cite{bengtzeliusDynamicsSupercooledLiquids1984,kirkpatrickScalingConceptsDynamics1989,debenedettiSupercooledLiquidsGlass2001,parisiMeanfieldTheoryHard2010,berthierTheoreticalPerspectiveGlass2011}.  
In most theoretical/computational studies, the liquids/glasses are treated classically which is justified given that
 many common glass-formers, such as silica, 
are composed of heavy atoms and exhibit a high glass transition temperature ($>1000$~K).  There are, however, 
liquids that vitrify at relatively low temperatures where nuclear quantum effects (due to the atoms delocalization) can play a relevant role.  
This is the case of small molecules that contain hydrogen, such as water ~\cite{eltarebPotentialEnergyLandscape2024a}, where
 isotope substitution effects are non-negligible even close to room temperature ~\cite{gainaruAnomalouslyLargeIsotope2014,giubertoniD2OImperfectReplacement2023}.  Obviously,
nuclear quantum effects (NQE) play a relevant role in light-element liquids, such as $\mathrm{H_2}$ and $\mathrm{He}$ (even at temperatures where 
classical statistics holds ~\cite{moralesEvidenceFirstorderLiquidliquid2010,kinugawaQuantumPolyamorphismCompressed2021,tsujimotoTwoLiquidStates2024}).  We also note that there are quantum systems,
 other than liquids, that exhibit a glass transition. These include electrons in metals \cite{amirSlowRelaxationsAging2009,mullerGlassTransitionCoulomb2004} and spin systems \cite{wuClassicalQuantumGlass1991,harrisPhaseTransitionsProgrammable2018,charbonneauSpinGlassTheory2023}. It remains unclear how quantum mechanics may change the behavior of low-temperature liquids close to their 
glass transition temperature, relative to the classical case. For example, an extension of mode coupling theory suggest that 
including quantum effects may increase the glass transition temperature of a liquid ~\cite{marklandQuantumFluctuationsCan2011}. In other systems, quantum effects may erase the glass state altogether ~\cite{klemmQuantumEffectsSpin1979}.

A well-established theoretical framework to study {\it classical} low-temperature liquids and glasses is the potential energy landscape (PEL) 
formalism ~\cite{stillingerHiddenStructureLiquids1982,stillingerEnergyLandscapesInherent2015}.  The PEL formalism is based on statistical mechanics~\cite{sciortinoPotentialEnergyLandscape2005, heuerExploringPotentialEnergy2008, walesExploringEnergyLandscapes2018} and has 
been applied extensively to describe the behavior of atomistic ~ \cite{sastrySignaturesDistinctDynamical1998,saika-voivodFreeEnergyConfigurational2004,zhouAnomalousPropertiesPotential2022} and molecular liquids ~\cite{starrThermodynamicStructuralAspects2001,lanaveLiquidStabilityModel2004}.  The PEL formalism provides (i) a qualitative description of classical 
liquids/glasses, and (ii) under simple approximations, it also provides the Helmholtz free energy and equation of state of the system ~\cite{robertsEquationStateEnergy1999,lanavePotentialEnergyLandscape2002,shellEnergyLandscapesIdeal2003,lanaveLiquidStabilityModel2004,handlePotentialEnergyLandscape2018,eltarebPotentialEnergyLandscape2024}.  In previous studies ~ \cite{giovambattistaPotentialEnergyLandscape2020,zhouHarmonicGaussianApproximations2024,eltarebPotentialEnergyLandscape2024a}, we extended the PEL formalism to the case of liquids that obey 
quantum mechanics.  It was shown that (i') the same qualitative description of classical
liquids and glasses, based on the PEL formalism, can be extended to quantum liquids. Moreover, path-integral molecular dynamics of water suggest that (ii') the PEL formalism may be used to predict the thermodynamic properties of liquids at low temperature in the presence of NQE ~\cite{eltarebPotentialEnergyLandscape2024a}.

In this work, we build upon the work presented in Refs.~\cite{giovambattistaPotentialEnergyLandscape2020,zhouHarmonicGaussianApproximations2024,eltarebPotentialEnergyLandscape2024a} and extend the definition of configurational entropy to the case of quantum liquids. The configurational entropy $S_{IS}$ ~\cite{stillingerHiddenStructureLiquids1982,sciortinoPotentialEnergyLandscape2005,stillingerEnergyLandscapesInherent2015},
as well as the associated Kauzmann temperature~\cite{kauzmannNatureGlassyState1948}, play important roles in the description of liquids and glasses in general ~\cite{debenedettiSupercooledLiquidsGlass2001,sastryRelationshipFragilityConfigurational2001,sciortinoPotentialEnergyLandscape2005, berthierConfigurationalEntropyGlassforming2019, parisiTheorySimpleGlasses2019}, and are essential to predict the thermodynamic behavior of low-temperature liquids, including the
corresponding equation of state.  To do so, in this work, we hypothesize that the distribution of local minima (inherent structures, IS) in the PEL of the $\it quantum$ liquid is identical to the distribution of IS in the PEL of the corresponding $\it classical$ liquid counterpart. This hypothesis, based on path-integral computer simulations of various model liquids, establishes a direct link between the configurational entropies of the quantum and associated classical liquid.
Using ring-polymer molecular dynamics (RPMD) simulations of a model liquid, we test that the 
configurational entropy defined in this study is consistent with relationships required by the PEL formalism.

It has been found that, for many classical liquids, the configurational entropy controls the corresponding liquid's dynamics.
Specifically, classical molecular dynamics (MD) simulations of different liquids, including Lennard-Jones binary mixtures (LJBM) ~\cite{sastrySignaturesDistinctDynamical1998}, silica ~\cite{saika-voivodFreeEnergyConfigurational2004}, 
and water \cite{scalaConfigurationalEntropyDiffusivity2000,handleAdamGibbsRelation2018}, show that the diffusion coefficient of the atoms/molecules in the system depends on the configurational entropy as predicted by the Adam-Gibbs (AG) relation ~\cite{adamTemperatureDependenceCooperative1965, handleAdamGibbsRelation2018, eltarebPotentialEnergyLandscape2024}. This implies 
that, at a given temperature, not only the thermodynamics but also the dynamics of the liquids are controlled by the PEL topography.  In this study, we also perform ring-polymer MD (RPMD) simulations and show that the AG relation, with the configurational entropy introduced here, holds for the model quantum liquid studied.

\section{Methods} 

In this work, we perform classical MD and RPMD 
simulations of the LJBM system defined in Ref.~\cite{kobTestingModecouplingTheory1995a}.
The LJBM model liquid is a good glass-former that has been studied extensively in the past using classical MD simulations~\cite{sastryRelationshipFragilityConfigurational2001,sciortinoExtensionFluctuationDissipationTheorem2001,berthierModernComputationalStudies2023}.
We consider a LJBM  composed of $N_A = 800$ type-A and $N_B = 200$ type-B particles interacting via Lennard-Jones pair
 interaction potentials $V_{\alpha\beta}(r)=4\epsilon_{\alpha,\beta}[ (\sigma_{\alpha\beta}/r)^{12} - (\sigma_{\alpha\beta}/r)^{6}]$ with  $\alpha\beta\in\{A,B\}$. Following Ref.~\cite{kobTestingModecouplingTheory1995a}, we use reduced units where $\sigma_{AA}=1.0, \sigma_{BB}=0.80,\sigma_{AB}=0.88$ (length), $\epsilon_{AA}=1.0,\epsilon_{BB}=0.5,\epsilon_{AB} = 1.5$ (energy), and $m_A=m_B = 1.0$ (mass); 
LJ pair interactions are truncated and shifted  at a cutoff distance $r_{c,\alpha\beta}=2.5~\sigma_{\alpha\beta}$, as implemented in the original work of Kob and Andersen~\cite{kobTestingModecouplingTheory1995a}.  Moreover, we also add a switching function to the pair interaction potentials, as implemented in OpenMM \cite{eastmanOpenMM7Rapid2017}, to make the forces smooth functions of the inter-particle distance $r$.  The switching function only affects the LJ pair interaction potential
at inter-particle distances  $r_{s,\alpha\beta}<r<r_{c,\alpha\beta}$ where $r_{s,\alpha\beta} = 0.9~r_{c,\alpha\beta}$.  The system is cubic  and periodic boundary conditions apply in all three directions.


As for the case of path-integral molecular dynamics (PIMD) simulations, the RPMD technique is based on the path integral formulation of statistical mechanics (see below) and they can be used to calculate thermodynamic (e.g., pressure) and structural properties (e.g., radial distribution functions) of a liquid in the presence of NQE. In PIMD/RPMD simulations, each atom of the system is represented by a ring-polymer composed of $n_b$ beads; by setting $n_b=1$, the PIMD/RPMD simulation technique reduces to classical MD simulations. To obtain $D(T)$, we use the RPMD technique which is based on RPMD simulations. As explained in Refs. \cite{craigQuantumStatisticsClassical2004,rossiHowRemoveSpurious2014}, from the RPMD trajectory, one can calculate, approximately, the dynamical properties of the quantum system.  In this work, $D(T)$ is calculated from the long-term behavior of the mean-square displacement of the ring-polymer centroids obtained from the RPMD simulations at constant temperature (see, e.g., \cite{marklandTheorySimulationsQuantum2012,eltarebPotentialEnergyLandscape2024a,millerIsomorphicClassicalMolecular2008}). 

To understand the differences in the PEL properties of the quantum and classical LJBM, we follow Refs.~\cite{giovambattistaPotentialEnergyLandscape2020, zhouHarmonicGaussianApproximations2024}
and study a family of LJBMs each characterized by a different value of the Planck's constant $h$.  
The quantum character of a liquid increases with increasing values of $h$ since the atom delocalization becomes 
more pronounced as $h$ increases (see below). 
In this study, we consider the cases $h_0 = 0.0000$, $h_a = 0.2474$, $h_b = 0.5000$, 
and $h_c = 0.7948$ (in reduced units of $\sigma_{AA}(\epsilon_{AA}m)^{1/2}$). The case $h = h_0$ 
corresponds to the classical liquid where the ring-polymers are collapsed at all times 
(for $h = 0$, the spring constant of the ring-polymers is $k^{sp} \to \infty$; see below). 
Hence, for the case $h = h_0$, we perform classical MD simulations. 
We note that the values of $h=h_a,h_b,h_c$ considered here are not negligible; for example, as 
discussed in detail in Ref.~\cite{zhouHarmonicGaussianApproximations2024} one obtains $h = 1.78$ and $h \approx 0.18$ (in reduced units) 
for the case of H$_2$ and argon, respectively. 

Most of the RPMD simulations are performed with $n_b = 10$ beads per ring-polymer. However, to test for $n_b$-effects,
 we also run RPMD simulations using $n_b = 20, 40 ~(h=h_c)$.  Classical MD simulations are run by setting $n_b=1$ in the RPMD simulations.
All computer simulations are performed at constant $(N, V, T)$ for the density $\rho = 1.2$ $(V/N = 9.4, N=1000$) and extend over a wide range of temperatures. Additional computer simulations at constant $(N,V,T)$  are performed at $T=5.0$ and over a wide range of volumes to calculate the total entropy of the system.
The temperature is controlled by using a local PILE thermostat with stochastic collision frequency $\gamma = 0.6$. 
At a given state point, the starting configuration of the RPMD simulation
 is taken from an independent equilibrium simulation performed at a very high temperature, $T=5.0$.
Equilibrium RPMD simulations extend for $t_{eq}\geq 100~\tau_{\alpha}$ where
 $\tau_{\alpha}$ is the relaxation time of the system defined 
as the time at which the mean-square displacement (MSD) of the type-A particles is $1.0 ~\sigma_{AA}^2$.
To obtain a large number of IS, we perform very long production runs, of $t \geq 1000 \tau_\alpha$ (see below).
All the RPMD simulations are performed using the OpenMM (version 7.3.0) software package~\cite{eastmanOpenMM7Rapid2017} modified to include the target Planck's constant $h$. 
In the OpenMM computer simulations, the atoms mass is set to
$m_A = 39.948 ~amu$, $\sigma_{AA} = 1.0$~\AA, and $\epsilon_{AA} = 1.0$~kJ/mol (but all quantities are reported in reduced 
units). The simulation time step for MD/RPMD is $dt= 0.01$~ps ($T\leq1.0$) and $dt=0.001$~ps ($T\in (1.0,5.0]$) for $h_0,h_a,h_b$, and $dt= 0.005$~ps ($T\leq 1.0$) and $dt=0.0005$~ps ($T\in (1.0,5.0]$)  for $h=h_c$.

From the long RPMD production runs, at a given temperature, we extract more than $1000$ independent configurations. 
Two configurations are considered to be independent if they are separated in time by, at least, $\delta t=\tau_{\alpha}$. 
For each of these configurations, we calculate the corresponding IS by minimizing the potential energy of the system using the L-BFGS-B algorithm (the potential energy minimization is performed using the {\it scipy.optimize} routine in Python ~\cite{virtanenSciPy10Fundamental2020} with an error tolerance of $10^{-12}$, with the forces and total potential energy obtained from the OpenMM software package using the  OpenCL platform with double precision). The IS energy is obtained directly from the minimization algorithm.
For each IS, we also calculate analytically the components of the mass-weighted Hessian matrix.
The mass-weighted Hessian matrix is then diagonalized to calculate the corresponding eigenvalues which provide the normal mode vibrational frequencies of the system (diagonalization is performed using the {\it Numpy} linear algebra package in Python  \cite{harrisArrayProgrammingNumPy2020}).


\section{The Potential Energy Landscape Formalism for Quantum Liquids}
\label{sec:PEL}
The PEL formalism provides a compact expression for the canonical partition function of the classical/quantum 
system of interest, $Q(N,V,T)$.  Next, we present a brief overview of the PEL formalism for a {\it quantum}
 liquid composed of $N$ atoms (not necessarily identical);  the generalization to molecular liquids is straightforward (see, e.g., Ref. \cite{eltarebPotentialEnergyLandscape2024a}).

{\it Defining a PEL for a quantum liquid}. 
For a quantum system, the canonical partition function is given by
\begin{eqnarray}  
Q(N,V,T) = \text{Tr}(\hat{\rho})
\label{partFunct-1}
\end{eqnarray}
where $\text{Tr}(\hat{\rho})$  is the trace of the density operator $\hat{\rho}=\exp(-\beta \hat{H})$  and
$\hat{H}$ is the Hamiltonian operator of the system,
\begin{eqnarray}  
\hat{H}=\sum_{i=1}^{N} \frac{\hat{\mathbf{p}}_i^2}{2m_i} + U(\hat{\mathbf{r}}_1,\hat{\mathbf{r}}_2,...,\hat{\mathbf{r}}_N)
\label{Hamilt}
\end{eqnarray}
In Eq.~\ref{Hamilt}, $U(\hat{\mathbf{r}}_1,\hat{\mathbf{r}}_2,...,\hat{\mathbf{r}}_N)$ is the potential energy operator, and  $(\hat{\mathbf{r}}_i, \hat{\mathbf{p}}_i)$ are  the position and momentum operators associated
with atom $i = 1, 2, ..., N$ ($\beta = 1/k_B T$ where $k_B$ is the Boltzmann's
constant).

Using the path-integral formulation of quantum statistical mechanics, it can be shown
 that the canonical partition function of the quantum liquid (Eq.~\ref{partFunct-1}) 
is {\it identical} to the canonical partition function of a {\it classical} system of $N$
(distinguishable) ring-polymers composed of $n_b \rightarrow \infty$ (distinguishable) 
beads~\cite{tuckermanStatisticalMechanicsTheory2010,ceperleyPathIntegralsTheory1995}. 
Specifically,
\begin{equation}  
    \begin{aligned}
        Q(N,V,T)= \lim_{n_b \rightarrow \infty} \frac{1}{h^{3n_bN}}\int_V \left(\prod_{i=1}^{N} \mathrm{d} \mathbf{r}_i^1  \cdots \mathrm{d} \mathbf{r}_i^{n_b} \right)\int_{-\infty}^{\infty} \left(\prod_{i=1}^{N} \mathrm{d} \mathbf{p}_i^1 \cdots \mathrm{d} \mathbf{p}_i^{n_b}\right) \\
        \exp\left(-\beta \mathcal{H}_{RP}( \mathbf{P}, \mathbf{R}) \right)
    \end{aligned} 
    \label{partFunct-2}
\end{equation}
where
\begin{eqnarray}  
\mathcal{H}_{RP}( \mathbf{P}, \mathbf{R} )= \sum_{i=1}^{N} \sum_{k=1}^{n_{b}} \frac{(\mathbf{p}_i^k)^{2}} {2m'_{i}}  +
                                          \sum_{i=1}^{N} \sum_{k=1}^{n_{b}} \frac{1}{2} k^{sp}_i (\mathbf{r}_i^{k+1}-\mathbf{r}_i^k )^{2} + \frac{1}{n_{b}} \sum_{k=1}^{n_{b}} U(\mathbf{r}_1^k,\mathbf{r}_2^k \ldots, \mathbf{r}_N^k) 
\label{Hrp}
\end{eqnarray}
is the Hamiltonian of the classical ring-polymer system.  For each atom $i=1,2,...,N$ of the quantum liquid
 with mass $m_i$, there is one and only one ring-polymer $i$ in
 the ring-polymer system associated to it.
The spring constant of the corresponding ring-polymer is given by $k^{sp}_i = \frac{m_i n_b}{(\hbar \beta)^2}$
 and the mass of the corresponding beads are given by $ m'_i = n_b m_i$. 
In Eq.~\ref{partFunct-2}, $(\mathbf{r}_i^k$,$\mathbf{p}_i^k)$  are the vector
position and momentum of the $k$-th bead of the $i$-th ring-polymer
($\mathbf{R}= \{ \mathbf{r}_i^{k} \}$ and $\mathbf{P}= \{ \mathbf{p}_i^{k} \}$; $i = 1, 2, ..., N$, $k = 1, 2, ..., n_b$). 
In Eq.~\ref{Hrp}, and throughout this work, $\mathbf{r}_i^1= \mathbf{r}_i^{n_b+1}$
for $i= 1, 2, ...,N$ since the polymers are ring-polymers.

Eq.~\ref{Hrp} implies that the potential energy of the ring-polymer system is given by
\begin{eqnarray}  
{\cal U}_{RP}(\mathbf{R}) = \sum_{i=1}^{N} \sum_{k=1}^{n_{b}} \frac{1}{2} k^{sp}_i (\mathbf{r}_i^{k+1}-\mathbf{r}_i^k )^{2} + \frac{1}{n_{b}} \sum_{k=1}^{n_{b}} U(\mathbf{r}_1^k,\mathbf{r}_2^k \ldots, \mathbf{r}_N^k).
\label{Urp}
\end{eqnarray}
As explained in Ref.~\cite{giovambattistaPotentialEnergyLandscape2020,zhouHarmonicGaussianApproximations2024}, the function 
${\cal U}_{RP}(\mathbf{R})$ defines a PEL that can be associated to the given quantum liquid.  
By applying the PEL formalism
(originally proposed to study classical liquids~\cite{stillingerHiddenStructureLiquids1982}) to the PEL defined by Eq.~\ref{Urp}, 
one can study the behavior of quantum liquids and, in particular,
 extract the corresponding thermodynamic properties~\cite{zhouHarmonicGaussianApproximations2024,eltarebPotentialEnergyLandscape2024a}.
We note that, strictly speaking, the above expressions hold for  $n_b \rightarrow \infty$.  
However, in path-integral computational studies, one chooses a sufficiently large value of
 $n_b$ for which the thermodynamic properties of the system of interest converge (i.e., they no
longer vary upon further increase in $n_b$). 

The PEL formalism provides a simple understanding for the behavior of low-temperature liquids and glasses.  Specifically,
within the PEL formalism, a classical/quantum liquid is represented by a point on the PEL that moves over time 
(in the case of the quantum liquid, such a 
PEL is given by Eq.~\ref{Urp} with a fix value of $n_b$).  At high temperatures, the liquid has sufficient 
kinetic energy to overcome the potential energy barriers of the PEL and hence, it can explore different basins of the PEL.  
In the liquid state, the molecules/atoms are able to diffuse with time and the system moves on the 
PEL describing a trajectory.  Instead, in the glass state, the molecules/atoms of the system are only able to 
vibrate about fix positions.  Accordingly, 
the representative point of the system in the PEL is limited to move within a single basin, about the corresponding local 
minimum or inherent structure (IS).   
While this picture holds for classical liquids and quantum liquids/ring-polymer systems, important differences exist in the
corresponding PEL formalism.  Among them is the fact that the PEL of a quantum liquid is $T$-dependent 
(since the spring constant $k^{sp}_i \propto T^2$) while the PEL of a classical system is not.

{\it Partition function in the PEL formalism.}  As explained in detail in Refs.~\cite{zhouHarmonicGaussianApproximations2024,eltarebPotentialEnergyLandscape2024a}, in the PEL formalism, 
the partition function of a quantum liquid (Eqs.~\ref{partFunct-1} and \ref{partFunct-2}) can be written as,
\begin{eqnarray}
Q(N, V, T) = \sum_{e_{IS}}  e^{-\beta(e_{IS}-T S_{IS}(N,V,T,e_{IS})+F_{vib}(N,V,T,e_{IS}))}
\label{Qdef}
\end{eqnarray}
where $S_{IS}(N, V, T, e_{IS})$ is the configurational entropy of the system and $F_{vib}(N, V, T, e_{IS})$ is the vibrational Helmholtz free energy of the system. Both quantities are precisely defined in the PEL formalism; see Refs. ~\cite{zhouHarmonicGaussianApproximations2024,eltarebPotentialEnergyLandscape2024a}. $S_{IS}(N, V, T, e_{IS})$ quantifies the number of IS available in the PEL with a given energy $e_{IS}$.  $F_{vib}(N, V, T, e_{IS})$ is the contribution to the Helmholtz free energy of the system, $F(N, V, T)$, due to the explorations (by the system) of the PEL basins with IS energy $e_{IS}$.

The sum in Eq.~\ref{Qdef} runs over all IS energies $e_{IS}$ available in the PEL.  In the thermodynamic limit, one may employ the saddle point approximation, so that  only the term that maximizes the sum in Eq.~\ref{Qdef} dominates. i.e., 

\begin{eqnarray}  
Q(N,V,T) \approx  
e^{-\beta \left( E_{IS} - T S_{IS}(N,V,T, E_{IS}) + F_{vib}(N,V,T, E_{IS}) \right) }
\label{Qfinal}
\end{eqnarray}
where $E_{IS}(N,V,T)$ is the solution to the following equation,
\begin{eqnarray}  
1-  T \left( \frac{\partial  S_{IS}(N,V,T, e_{IS})}{\partial e_{IS}} \right)_{N,V,T}  +
 \left( \frac{\partial F_{vib}(N,V,T, e_{IS}) }{\partial e_{IS}}   \right)_{N,V,T} =0
\label{EisDef}
\end{eqnarray}
In computational studies, one identifies $E_{IS}(N,V,T)$ 
with the average value of $e_{IS}$ sampled by the system at the given 
working conditions $(N,V,T)$.

\subsection{Gaussian and Harmonic PEL}

A common approximation in PEL studies is to assume that the PEL is Gaussian ~\cite{sastryRelationshipFragilityConfigurational2001,stillingerEnergyLandscapesInherent2015,sciortinoPotentialEnergyLandscape2005},
i.e., that the distribution of IS with energy $e_{IS}$ available in the PEL is given by a Gaussian distribution
\begin{eqnarray}  
\Omega_{IS}(N, V,T, e_{IS})\approx \frac{1}{\sqrt{2 \pi} \sigma} e^{\alpha N} e^{-(e_{IS}-E_0)^2/2 \sigma^2}
\label{GaussPEL}
\end{eqnarray}
This implies that the configurational entropy of the system is given by
\begin{eqnarray}    
S_{IS}(N,V,T, e_{IS}) \approx  k_B \left[ \alpha N - \frac{(e_{IS}-E_0)^2}{2 \sigma^2}     \right]
\label{SconfGauss}
\end{eqnarray}
Here, $\alpha$, $E_0$, and $\sigma$ are PEL variables that depend on $(V,T)$.
Importantly, MD/PIMD simulations of very different liquids, including LJBM~\cite{sastrySignaturesDistinctDynamical1998,heuerWhyDensityInherent2000}, water~\cite{handlePotentialEnergyLandscape2018,eltarebPotentialEnergyLandscape2024a,scalaConfigurationalEntropyDiffusivity2000}, ortho-terphenyl~\cite{lanavePotentialEnergyLandscape2002}, and water-like monatomic systems~\cite{neophytouPotentialEnergyLandscape2024,zhouHarmonicGaussianApproximations2024}, indicate that the corresponding PEL is Gaussian. 

Another common approximation in the PEL formalism is to assume that the PEL basins are quadratic functions
(of the atoms/ring-polymer beads coordinates) near the corresponding IS.
Under the harmonic approximation of the PEL, one can show that
\begin{equation}    
\begin{aligned}
F_{vib}(N,V,T, e_{IS}) \approx F^{harm}_{vib}(N,V,T, e_{IS}) = 3N n_b k_B T \ln \left( \beta \hbar \omega_0 \right) + \\
k_B T  {\cal S}(N,V,T, e_{IS})
\end{aligned}
\label{FvibHA}
\end{equation}
where 
\begin{eqnarray}    
{\cal S}(N,V,T, e_{IS}) \approx  \left<  \ln \left( \prod_{j=1}^{3n_bN} \left( \omega_j/ \omega_0 \right) \right) \right>_{e_{IS}} 
\label{ShapeFunctionbDef}
\end{eqnarray}
is the basin-shape function.
The $3 n_b N$ values $\{ \omega_j^2= \omega_j^2(N,V,T,e_{IS}) \}$ are the eigenvalues of the {\it mass-weighted} Hessian matrix of the ring-polymer system evaluated 
at the IS with energy $e_{IS}$; $\left< ... \right>_{e_{IS}}$
indicates an average over all basins of the PEL with energy $e_{IS}$. 
The constant $\omega_0$ is an arbitrary quantity that makes the argument of $\ln(...)$ dimensionless.
${\cal S}(N,V,T, e_{IS})$ quantifies the average local curvature of the PEL basins with IS
energy $e_{IS}$ and it is the only term in Eq.~\ref{FvibHA} that makes $F_{vib}$
dependent on the PEL of the system. We note that Eqs.~\ref{Qfinal}-\ref{ShapeFunctionbDef} also apply to the case of classical liquids.  However, since 
the PEL of classical liquids is $T$-independent, the topographic properties of the PEL are also $T$-independent;
 specifically,  $S_{IS}= S_{IS}(N,V,e_{IS})$, 
 $\omega_j = \omega_j(N, V, e_{IS})$ and ${\cal S}= {\cal S}(N,V, e_{IS})$.

In the PEL formalism, the Helmholtz free energy of the system, $F(N,V,T)= -k_B T \ln(Q(N,V,T))$, follows directly from Eq.~\ref{Qfinal},
\begin{eqnarray}    
F(N,V,T) =  E_{IS}(N,V,T)-T S_{IS}(N,V,T,E_{IS}) + F_{vib}(N,V,T,E_{IS})
\label{F-PEL}
\end{eqnarray}
In the case of a Gaussian and Harmonic approximation of the PEL (Eqs.~\ref{SconfGauss} and \ref{FvibHA}),
it can be shown from Eq.~\ref{F-PEL} that the free energy of the quantum liquid can be expressed in 
terms of only three PEL variables $\{ \alpha, E_0, \sigma \}$, and the eigenvalues of the mass-weighted Hessian matrix of the corresponding ring-polymer system; see, e.g., 
Refs.~\cite{sciortinoPotentialEnergyLandscape2005,heuerExploringPotentialEnergy2008,stillingerEnergyLandscapesInherent2015,zhouHarmonicGaussianApproximations2024}.
Accordingly, one can obtain an analytical expression for all the thermodynamic properties of the quantum liquid, including the equation of 
state~\cite{robertsEquationStateEnergy1999,handlePotentialEnergyLandscape2018,neophytouPotentialEnergyLandscape2024,eltarebPotentialEnergyLandscape2024}. 
 In particular, it can be shown that, for a Gaussian and harmonic PEL, the total energy of the system,  
$E(N,V,T) = ( \partial (\beta F)/ \partial \beta)_{N,V}$, is given by (see Appendix \ref{AppendixA})
\begin{eqnarray}
E(N,V,T)= E_{IS}(N,V,T) +  E_{vib}(N,V,T)
\label{Evib_def}
\end{eqnarray}
where 
\begin{eqnarray}
E_{IS}(N,V,T) \approx E^{harm}_{IS}(N,V,T) = E_0(V,T) - \sigma^2(V,T) (\beta + b)
\label{Eis_GAHA}
\end{eqnarray}
and
\begin{eqnarray}    
E_{vib}(N,V,T) \approx E^{harm}_{vib}(N,V,T) = 3 N n_b k_B T 
		+ \left( \frac{\partial  {\cal S}}{\partial \beta}   \right)_{N,V,E_{IS}}
\label{Evib_GAHA}
\end{eqnarray}
In Eq.~\ref{Eis_GAHA}, $b \equiv \left(\frac{ \partial {\cal S}(N,V,T, e_{IS})}{\partial e_{IS}} \right)_{N,V,T,e_{IS}=E_{IS}}$. As explained in Ref.~\cite{zhouHarmonicGaussianApproximations2024}, to obtain Eqs.~\ref{Eis_GAHA} and \ref{Evib_GAHA}, we have assumed that 
the parameters $\alpha=\alpha(V)$, $E_0=E_0(V)$, and $\sigma=\sigma(V)$, i.e., they are $T$-independent. We note that Eq.~\ref{Eis_GAHA} and~\ref{Evib_GAHA} are also valid for classical liquids. In the classical case, however, Eq.~\ref{Evib_GAHA} reduces to $E_{vib}=3NkT$ since $n_b=1$ and the last term of Eq.~\ref{Evib_GAHA} vanishes (the PEL for classical system is $T$-independent).  Alternatively, it can be shown that $\left(\frac{\partial {\cal S}(N,V,T,e_{IS})}{\partial \beta}\right)_{N,V,e_{IS}=E_{IS}}= E_{sp}(N,V,T)$ [see Appendix~\ref{AppendixB}] where $E_{sp}(N,V,T)\equiv\langle \sum_i \sum_k^{n_b}  \frac{1}{2} k^{sp}_i (\mathbf{r}_i^{k+1} - \mathbf{r}_i^k)^2\rangle_{N,V,T}$ is the potential energy of the ring-polymer {\it springs}. For classical systems, $n_b=1$, (there are no springs associated to the liquid atoms) and hence, $E_{sp}(N,V,T)=0$.

Relevant to this work, we note that Eqs.~\ref{Eis_GAHA} and ~\ref{Evib_GAHA} hold for the case 
of {\it classical} LJBM~\cite{sciortinoInherentStructureEntropy1999,sciortinoThermodynamicsSupercooledLiquids2000} suggesting that its PEL is indeed Gaussian and harmonic. 
In the classical case, $E_{vib}= 3Nk_B T$ since the PEL and hence, ${\cal S}$, are $T$-independent.  
Below we show that Eqs.~\ref{Eis_GAHA} and ~\ref{Evib_GAHA}
also hold for quantum LJBM if the Planck's constant is small (mild quantumness, $h=h_a$)
 but fail as the Planck's constant is increased. 
For large values of $h$, anharmonic correction to the harmonic approximation of the PEL are needed.

\subsection{Anharmonic Corrections in the PEL Formalism}

Anharmonic corrections to the PEL have been introduced in the past to study the 
equation of state of {\it classical} liquids using the PEL formalism~\cite{lanavePotentialEnergyLandscape2002,sciortinoPotentialEnergyLandscape2005,handlePotentialEnergyLandscape2018,eltarebPotentialEnergyLandscape2024}. In most studies, 
anharmonicities were assumed to depend only on $T$, implying that the basins 
anharmonicities were independent of where the basins were located within the PEL (the basins anharmonicities were assumed to be independent of the basin IS energy $e_{IS}$). 
 Here, we present a general approach where the anharmonic corrections to the PEL may vary with both $T$ and $e_{IS}$. 

In the presence of anharmonic corrections, one can \textit{formally} express the vibrational Helmholtz free energy of the system as 
\begin{equation}
    F_{vib}(N,V,T) = F^{harm}_{vib}(N,V,T) + F^{anh}_{vib}(N,V,T,E_{IS}),
    \label{FvibAnh}
\end{equation}
where
$F^{harm}_{vib}(N,V,T)$ is given by Eq.~\ref{FvibHA} and $F^{anh}_{vib}(N,V,T,E_{IS})$ is 
the correction due to the basins anharmonicity. The formal expression for the total energy of the system, 
$E(N,V,T) = (\partial \beta F(N,V,T)/ \partial \beta)_{N,V}$,
follows from Eqs.~\ref{F-PEL} and \ref{FvibAnh} .   It can be shown that $E(N,V,T)$ is also given by Eqn. \ref{Evib_def}, however, the new expressions for $E_{IS}$ and  $E_{vib}$ (for a Gaussian PEL with anharmonic corrections) change as follows (see Appendix~\ref{AppendixA}).
\begin{eqnarray}
E_{IS}(N,V,T) = E^{harm}_{IS}(N,V,T) + E^{anh}_{IS}(N,V,T)
\label{Eis_GAANH}
\end{eqnarray}
where $ E^{harm}_{IS}(N,V,T)$ is given by Eq.~\ref{Eis_GAHA}, and
\begin{eqnarray}
E^{anh}_{IS}(N,V,T)= - \sigma^2  \left( \frac{ \partial (\beta F^{anh}_{vib}(N,V,T,e_{IS})) } { \partial e_{IS} }  \right)_{N,V,T,e_{IS}= E_{IS}}.
\label{EisAnh-GAANH}
\end{eqnarray}
The new expression for $E_{vib}$  is given by
\begin{eqnarray}
E_{vib}(N,V,T) = E^{harm}_{vib}(N,V,T) + E^{anh}_{vib}(N,V,T, E_{IS}) 
\label{Evib_GAANH}
\end{eqnarray}
where $E^{harm}_{vib}(N,V,T)$ is given by Eq.~\ref{Evib_GAHA}, and
\begin{eqnarray}  
E^{anh}_{vib}(N,V,T)= \left(  \frac{  \partial (\beta F^{anh}_{vib}(N,V,T, E_{IS})) }  { \partial \beta }  \right)_{N,V,E_{IS}}. 
\label{EvibAnh-GAANH}
\end{eqnarray}

Eqs.~\ref{EisAnh-GAANH} and \ref{EvibAnh-GAANH} are formal expressions that yield $E_{IS}^{anh}$ and $E_{vib}^{anh}$ once $F_{vib}^{anh}$ is provided. As an example, we consider the case where the anharmonic corrections depend only on $T$. In this case, one can write the following general expression, \cite{sciortinoPotentialEnergyLandscape2005}
\begin{eqnarray}
F^{anh}_{vib}(N,V,T,E_{IS}) = \sum_{i=2}^{i_{max}} \frac{c_i}{1-i} T^i,
\label{FanhT}
\end{eqnarray}
where the coefficients $\{ c_i \}$ depend only on $(V,N)$. In this case, Eq.~\ref{EvibAnh-GAANH} implies that 
\begin{eqnarray}
E^{anh}_{vib}(N,V,T)= \sum_{i=2}^{i_{max}} c_i  T^i
\label{EvibanhT}
\end{eqnarray}
while $E_{IS}= E^{harm}_{IS}$ (from Eq.~\ref{EisAnh-GAANH}, $E^{anh}_{IS}=0$).
Eqs.~\ref{FanhT} and \ref{EvibanhT}  are consistent with Refs.~\cite{handlePotentialEnergyLandscape2018,eltarebPotentialEnergyLandscape2024}.

In this work, we will model the anharmonicities of the PEL basins of the quantum LJBM as $\beta F_{vib}^{anh}= \sum_{i=1}^{\infty} \tilde{B_i}(N,V,T) e_{IS}^i$ where the coefficients $\tilde{B_i}(N, V, T)$ can obtained from the RPMD simulations (see below).  However, we find that keeping this expansion up to first order in $e_{IS}$ is sufficient to fit the LJBMs studied in this work, 
\begin{eqnarray}
\beta F^{anh}_{vib}(N,V,T,e_{IS}) = \tilde{B}_0(N,V,T) + \tilde{B}_1(N,V,T)~e_{IS}
\label{Fanh_expli}
\end{eqnarray}
Using Eqs.~\ref{EisAnh-GAANH} and \ref{EvibAnh-GAANH}, one obtains
\begin{eqnarray}
    E_{IS}^{anh}(N, V, T) = - \sigma^2 \tilde{B}_1(N, V, T)
\label{Eis_GAANH_expli}
\end{eqnarray}
and
\begin{eqnarray}
E_{vib}^{anh}(N,V,T, E_{IS})= \left(  \frac{  \partial \tilde{B}_0(N,V,T)  }  { \partial \beta }  \right)_{N,V} + \left(  \frac{  \partial \tilde{B}_1(N,V,T)  }  { \partial \beta }  \right)_{N,V}  E_{IS}(N,V,T)
\label{Evib_GAANH_expli}
\end{eqnarray}

To summarize, for a Gaussian PEL with anharmonicities given by Eq.~\ref{Fanh_expli}, the following expressions hold (using Eqs.~\ref{FvibAnh},\ref{Eis_GAANH},\ref{Evib_GAANH},\ref{Fanh_expli},\ref{Eis_GAANH_expli},\ref{Evib_GAANH_expli}),
\begin{equation}
\begin{aligned}
    F_{vib}(N,V,T,e_{IS})= 3N n_b k_B T \ln \left( \beta \hbar \omega_0 \right)
                       + k_B T  {\cal S}(N,V,T, e_{IS})    \\
                       + \tilde{B}_0(N,V,T) + \tilde{B}_1(N,V,T)~e_{IS} 
\label{FvibFinal}
\end{aligned}
\end{equation}
\begin{eqnarray}
E_{IS}(N,V,T) = E_0(V)-\sigma(V)^2 \left(\beta +b(N,V,T)+\tilde{B_1}(N,V,T)\right)
\label{EisFinal}
\end{eqnarray}
\begin{equation}
\begin{aligned}
E_{vib}(N,V,T) = 3N n_b k_B T 
+ \left( \frac{\partial {\cal S}(N, V, T,E_{IS})}{\partial \beta} \right)_{N,V, E_{IS}} 
+ \left( \frac{\partial {\tilde B}_0(N,V,T)}{\partial \beta} \right)_{N,V} \\
+ \left( \frac{\partial {\tilde B}_1(N,V,T)}{\partial \beta} \right)_{N,V}  E_{IS}(N,V,T)
\label{EvibFinal}
\end{aligned}
\end{equation}

\subsection{Hypotheses and Protocols}
\label{sec:hypotheses}
The theoretical predictions of the PEL formalism (Sec.~\ref{sec:PEL}) are based on a handful of PEL variables, $\{ \alpha, E_0, \sigma^2, b, {\cal S}, \tilde{B}_0, \tilde{B}_1 \}$. For any practical purpose, these quantities need to be fit to properties obtained from the RPMD simulations.   Next, we explain how these quantities are obtained, and the approximations involved. 

{\it Quantities $\{ {\cal S}, b \}$ }.
For the classical LJBM, the normal mode frequencies $\{ \omega_{j,0} \}$ 
with $j=1,2,...,3N$ are calculated by diagonalizing the 
corresponding {\it mass-weighted} Hessian matrix evaluated at the IS sampled by the system; the eigenvalues of the {\it mass-weighted} Hessian matrix 
are $\{ (\omega_{j,0} )^2 \}$.  For the quantum liquid/ring-polymer systems, there are 
$3 n_b N$ normal mode frequencies $\{ \omega_j \}$ ($j=1,2,..., 3 n_b N$)
 given by the eigenvalues of the {\it mass-weighted} Hessian matrix
of the ring-polymer system evaluated at the corresponding IS [Eq.~\ref{ShapeFunctionbDef}].  In this case, we follow the procedure described in Refs.~\cite{zhouHarmonicGaussianApproximations2024,eltarebPotentialEnergyLandscape2024a} to
obtain the ring-polymer (IS) normal mode frequencies analytically, using the values $\{ \omega_{j,0} \}$ of the corresponding classical system; 
see Eq.~20 in Ref.~\cite{eltarebPotentialEnergyLandscape2024a}.

The normal mode frequencies so obtained are then used in Eq.~\ref{ShapeFunctionbDef} to calculate the
 shape function of the system.  As shown in the supplementary information (SI), and consistent with previous MD/PIMD studies of 
classical and quantum liquids, our path-integral computer simulations of the LJBM show that
\begin{eqnarray}
{\cal S}(N,V,T,e_{IS}) \approx a(N,V,T) +  b(N,V,T)~ e_{IS}
\label{shape-b}
\end{eqnarray}
for all values of $h$ studied. Note that for quantum liquids, the variables $a$ and $b$ depend on $(N,V,T)$ while, in the classical case, they depend only on $(N,V)$.

{\it Quantities $\{ \alpha, E_0, \sigma^2 \}$}. These parameters characterize the configurational entropy of the system
 $S_{IS}(N,V,T,e_{IS})$ and hence, they specify how the IS are distributed within the PEL (Eqs.~\ref{GaussPEL} and \ref{SconfGauss}). 
In our previous studies~\cite{zhouHarmonicGaussianApproximations2024,eltarebPotentialEnergyLandscape2024a}, we assumed that, for the 
quantum liquids, $\{ \alpha, E_0, \sigma^2 \}$ were independent of $T$.  This implies that, while the PEL of the quantum 
liquid is $T$-dependent, the distribution of IS within the PEL is not. 

In previous studies \cite{giovambattistaPotentialEnergyLandscape2020,zhouHarmonicGaussianApproximations2024,eltarebPotentialEnergyLandscape2024a}, we found that the Gaussian approximation of the PEL works remarkably well for atomistic and molecular quantum liquids.
In the absence of anharmonicities, Eq.~\ref{Eis_GAHA} was found to be in very good agreement with the
values of $E_{IS}(T)$ obtained from RPMD simulations of an atomistic and molecular systems \cite{zhouHarmonicGaussianApproximations2024,eltarebPotentialEnergyLandscape2024a}. Importantly, (i) it was found that the parameters  $\{ \alpha, E_0, \sigma^2 \}$ (in the absence of anharmonicities) depend on the quantumness of the liquid (i.e., they all depend on $h$).  
 
This is not unreasonable since, varying $h$, changes the Hamiltonian of the ring-polymer system (Eq.~\ref{Hrp}). However, we also found indications that the values of $\{ \alpha, E_0, \sigma^2 \}$ should {\it not} depend on the nature (classical/quantum) 
of the liquid (as quantified by $h$).  Specifically, if the values of $\{ \alpha, E_0, \sigma^2 \}$ vary with $h$
then the distributions of the IS in the PEL of classical and quantum liquids must be different. For this to be the case, 
there should be IS in the ring-polymer system PEL (RP-PEL) where the ring-polymers are {\it not} collapsed  -- 
as explained in Ref.~\cite{zhouHarmonicGaussianApproximations2024}, if the ring-polymers are collapsed at the IS of the RP-PEL then such IS (in the RP-PEL) also define an IS of the classical liquid PEL (CL-PEL), and {\it vice versa} ~\cite{zhouHarmonicGaussianApproximations2024}.  
However, (ii) for all the quantum liquids studied~\cite{giovambattistaPotentialEnergyLandscape2020,zhouHarmonicGaussianApproximations2024,eltarebPotentialEnergyLandscape2024a}, it was always found that the 
ring-polymers were collapsed at the IS of the RP-PEL sampled by the system. 

There are two options for the findings (i) and (ii) to be compatible. (A) One may assume that the quantities 
$\{ \alpha, E_0, \sigma^2 \}$ vary with the quantumness of the liquid {\it but} the IS of the RP-PEL with {\it non-collapsed} ring-polymers are rare, difficult to sample in the RPMD simulations. Alternatively, one may consider that (B) there are no IS 
in the RP-PEL where ring-polymers are not collapsed at the working conditions and hence,  
the quantities $\{ \alpha, E_0, \sigma^2 \}$ are independent of whether the liquid obeys classical or quantum mechanics. 
In this work, we will assume that option (B) holds.   Accordingly, the 
values of $\{ \alpha, E_0, \sigma^2 \}$ are considered to be $h$-independent and hence, they will be extracted from the classical MD simulations, following the same procedure from previous classical MD simulations  ~\cite{sciortinoThermodynamicsSupercooledLiquids2000,sastryRelationshipFragilityConfigurational2001,scalaConfigurationalEntropyDiffusivity2000,handlePotentialEnergyLandscape2018,eltarebPotentialEnergyLandscape2024}. It follows from Eq.~\ref{SconfGauss}, that hypothesis (B) also implies that the configurational entropy $S_{IS}$ of the system as a function of $e_{IS}$ is identical for the classical and quantum liquid. As we will show, in the scenario (B), anharmonicities may need to be included.  

{\it Quantities $\{ \tilde{B}_0, \tilde{B}_1 \}$.}   
As a general expressions, we will consider that, for fix value of $(N,V)$,   
\begin{eqnarray}
\tilde{B}_j(T)= \sum_{i=0}^{i_{max}} c_{j,i} T^i
\label{Bj}
\end{eqnarray}
 where $j=0,1$. We find that a value of $i_{max}=3$ is sufficient to reproduce
 the results from the RPMD simulations. As will be shown below, the coefficients $c_{j=1,i}$ can be obtained from the RPMD simulations using Eq.~\ref{Eis_GAANH_expli}. Similarly, the coefficient  $c_{j=0,i}$ can be obtained from the RPMD simulations using Eq.~\ref{Evib_GAANH_expli} (the constant $c_{0,0}$ cannot be obtained in this way but plays no role in the configurational entropy $S_{IS}$, since it is an additive constant in the total free energy of the system).

In order to validate hypothesis (B), in Sec.~\ref{sec:PELBMLJ} we will test whether the configurational entropy given in Eq.~\ref{SconfGauss}, with the parameters $\{\alpha,E_0,\sigma^2\}$ of the classical LJBM, also hold for the quantum LJBMs studied. To do so, we note that the probability for the system to sample a basin 
with IS energy $e_{IS}$, at a given $T$ ($N,V$ are constant) is given by 
\begin{eqnarray}
\mathrm{P}(e_{IS}, T)= \frac{ e^{-\beta \left( e_{IS} - T S_{IS}(N,V,T, e_{IS}) + F_{vib}(N,V,T, e_{IS}) \right) } }{Q(N,V,T)}
\end{eqnarray}
Accordingly, $S_{IS}(N,V,T,e_{IS})$ must obey the following expression (see Eqs.~\ref{FvibHA} and \ref{FvibAnh}),
\begin{equation}
\begin{aligned}
S_{IS}(N,V,T,e_{IS})/k_B &= \ln \left( \mathrm{P}(e_{IS}, T) \right) + 3 N n_b \ln(\beta \hbar \omega_0) \\
&+ {\cal S}(N,V,T,e_{IS}) + \beta e_{IS} + \beta F^{anh}_{vib}(N,V,T, e_{IS}) + \ln(Q(N,V,T))  
\end{aligned}
\label{logPeis}
\end{equation}
In this work, we will calculate $\mathrm{P}(e_{IS}, T)$ directly from RPMD simulations of LJBMs and use Eq.~\ref{logPeis} to validate the expression for $S_{IS}(N,V,T,e_{IS})$ obtained from the PEL formalism (with the hypothesis (B) discussed above).

\section{Potential Energy Landscape of {\it classical} and {\it quantum} Lennard-Jones Binary Mixtures}
\label{sec:PELBMLJ}
\begin{figure}[!htb]
\centering{
\includegraphics[width=0.45\textwidth]{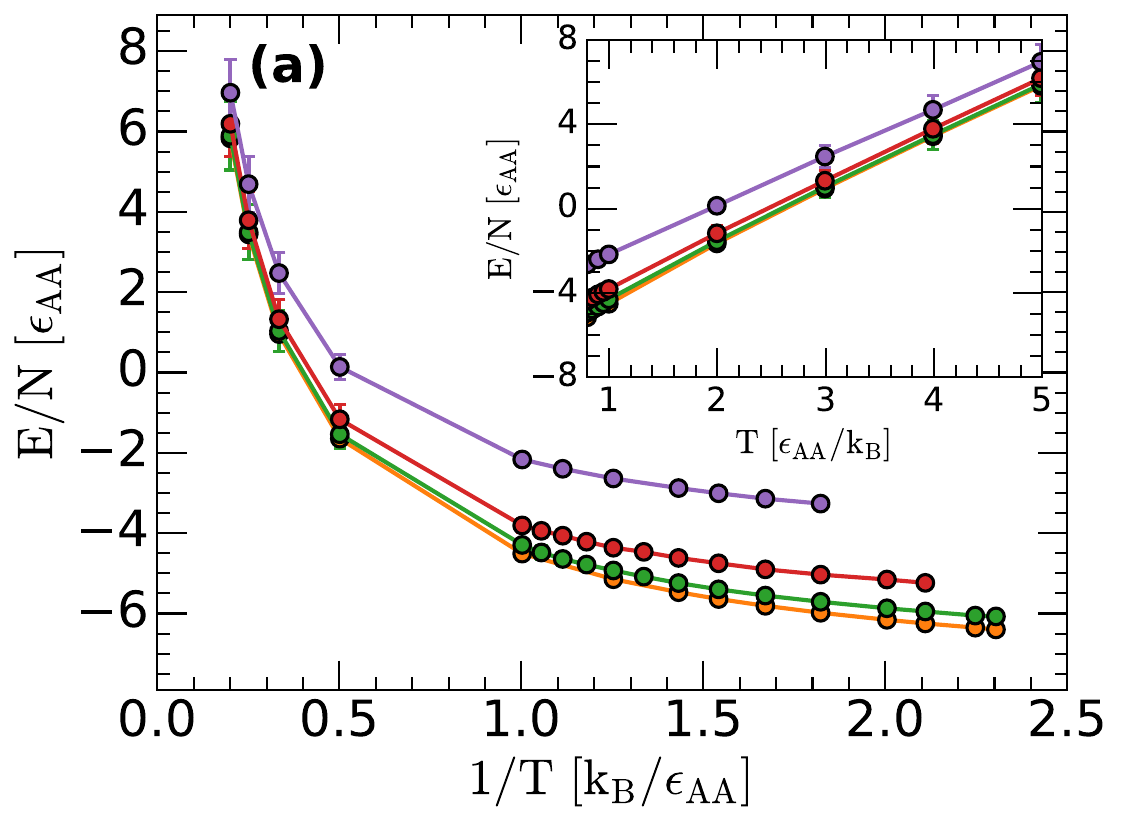}  
\includegraphics[width=0.45\textwidth]{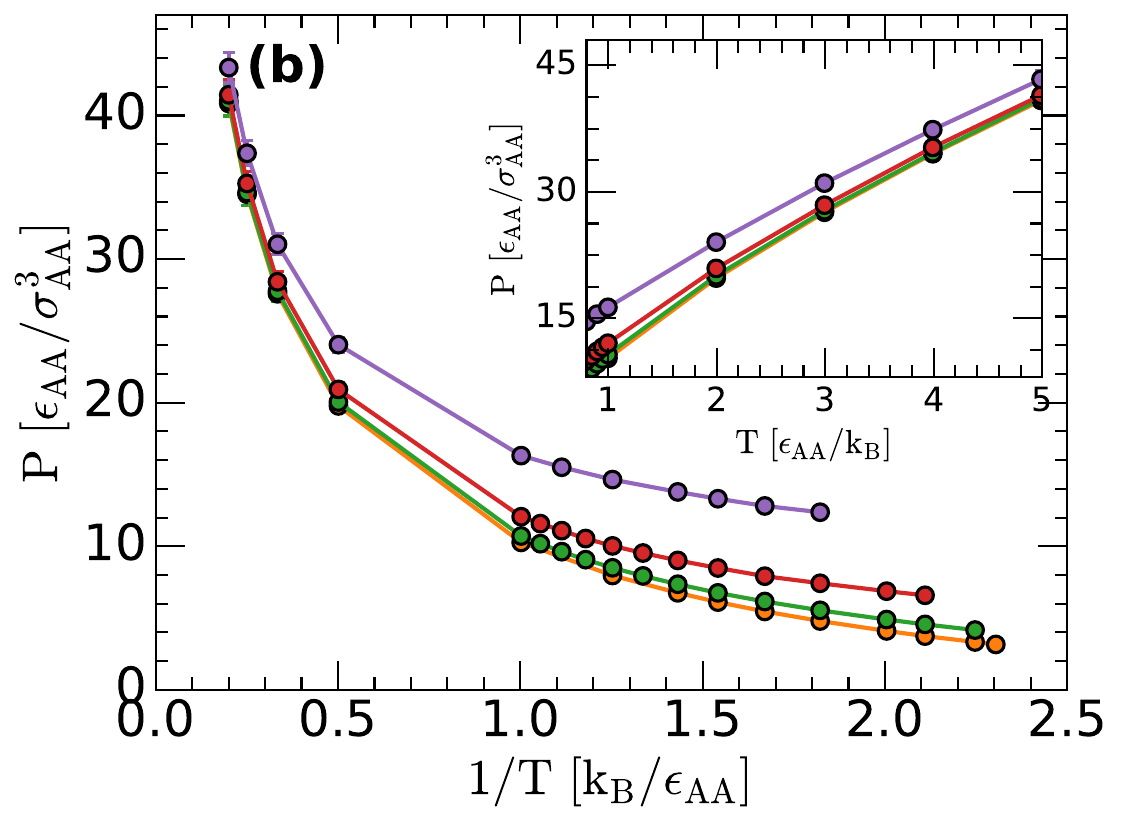}
\includegraphics[width=0.45\textwidth]{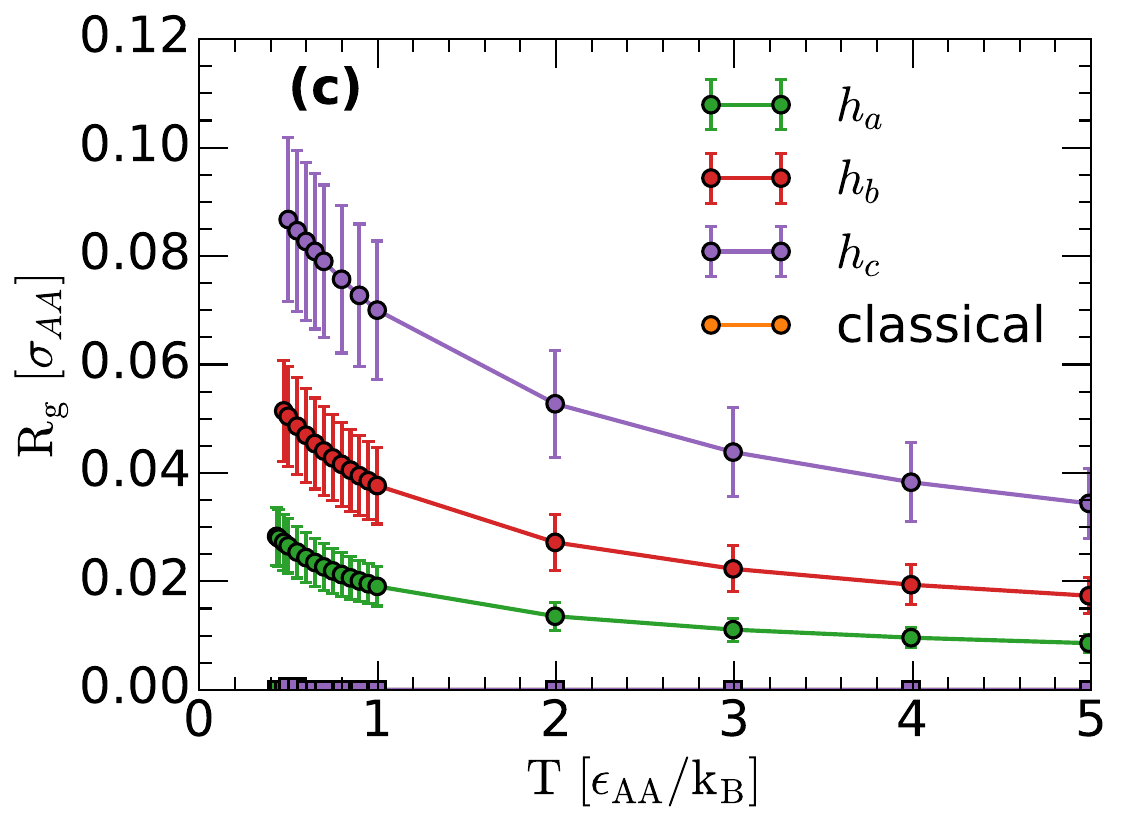}  
}
\caption{(a) Total energy and (b) pressure of the LJBMs as a function of temperature ($\rho=1.2$) obtained from RPMD simulations with Planck's constant $h=0$ (classical), $h_a=0.2474, h_b=0.5$, and $h_c=0.7948$.  As $h$ increases, and the LJBMs become more quantum, both $E(T)$ and $P(T)$ increase, particularly at low temperatures. (c) Radius of gyration $R_g$ of the A-type particles of the LJBMs studied (circles).  In all cases, $R_g(T)$ increases upon cooling.  The observed atoms delocalization becomes more pronounced with increasing $h$.  The squares in (c) correspond to the values of $R_g$ at the IS sampled by the system; $R_g(T) \approx 0$ at the IS implying that the ring-polymers are collapsed.}
\label{fig:E_P_sim}
\end{figure}

{\it Equilibrium properties}. We focus on the total energy $E(T)$ and pressure $P(T)$ of the 
LJBMs with different levels of quantumness. Figs.~\ref{fig:E_P_sim}(a) and \ref{fig:E_P_sim}(b) show $E(T)$ and $P(T)$ for the LJBMs obtained from MD/PIMD simulations with $h=0$ (classical) and $h = h_a, h_b, h_c$  (quantum). At high temperatures the values of $E(T)$ and $P(T)$ obtained for different values of $h$ become closer and closer with increasing temperatures.  Hence, quantum effects do not affect the thermodynamic properties of the LJBMs at very high temperatures (as expected). Instead, at low temperatures ($T < 1.0$), both $E(T)$ and $P(T)$ increase with increasing $h$, as the LJBMs become more quantum. This is due to the quantum delocalization of the atoms which becomes more pronounced with increasing $h$. Indeed, as shown in Fig.~\ref{fig:E_P_sim}(c), the radius of gyration, $R_g(T)$, of the ring-polymers associated to the LJBMs increases with increasing $h$.
We note that the atoms delocalization in the LJBMs studied is mild but not negligible.  At the lowest temperatures studied, $R_g \lesssim 0.09$; a value of $R_g =0.09$ indicates that the ring-polymers expand over a sphere of radius approximately equal to $9 \%$ of the A-particle hard-core radius.

\begin{figure}[!htb]
\centering{
    \includegraphics[width=0.45\textwidth]{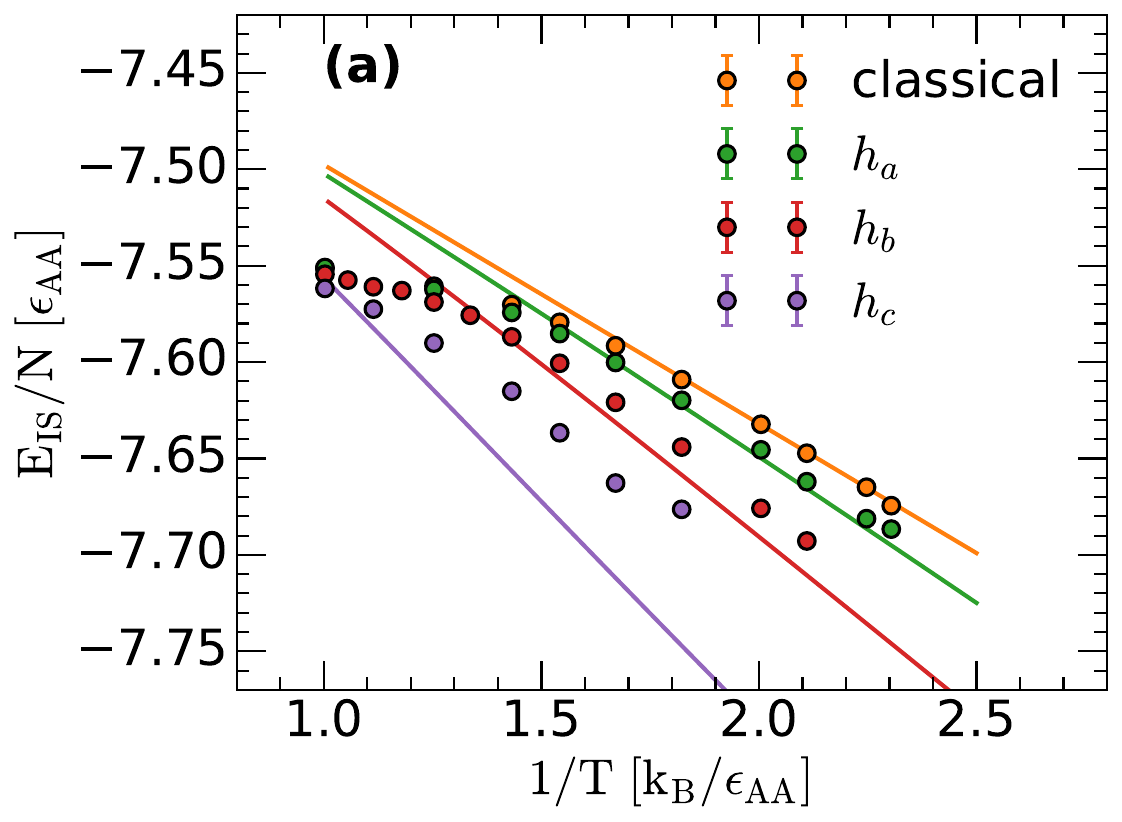}  
    \includegraphics[width=0.45\textwidth]{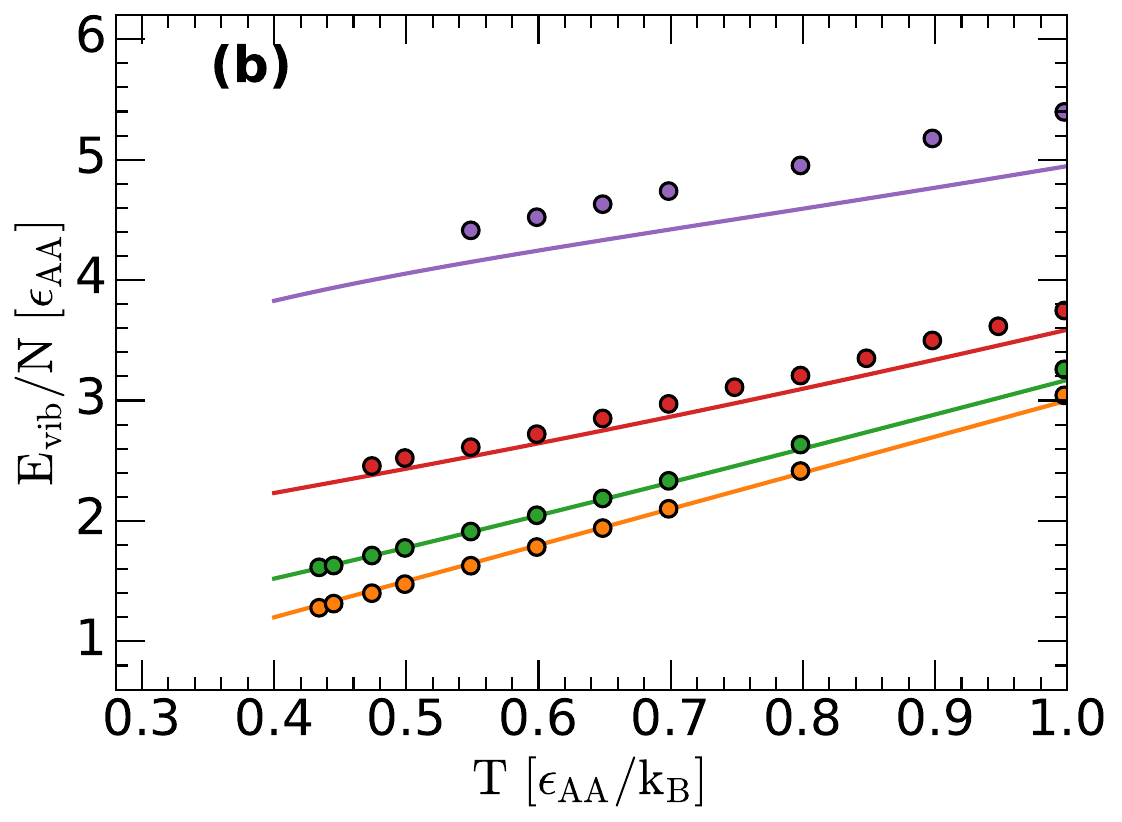}  
    \includegraphics[width=0.45\textwidth]{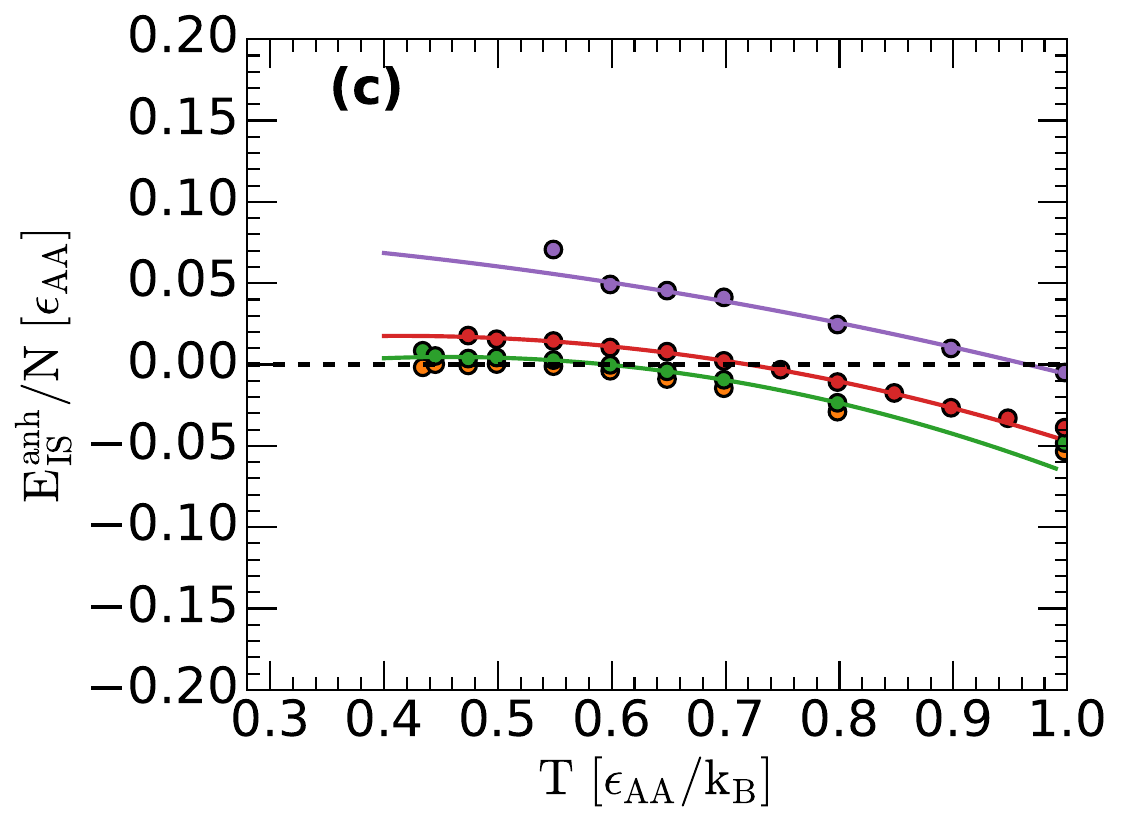}  
    \includegraphics[width=0.45\textwidth]{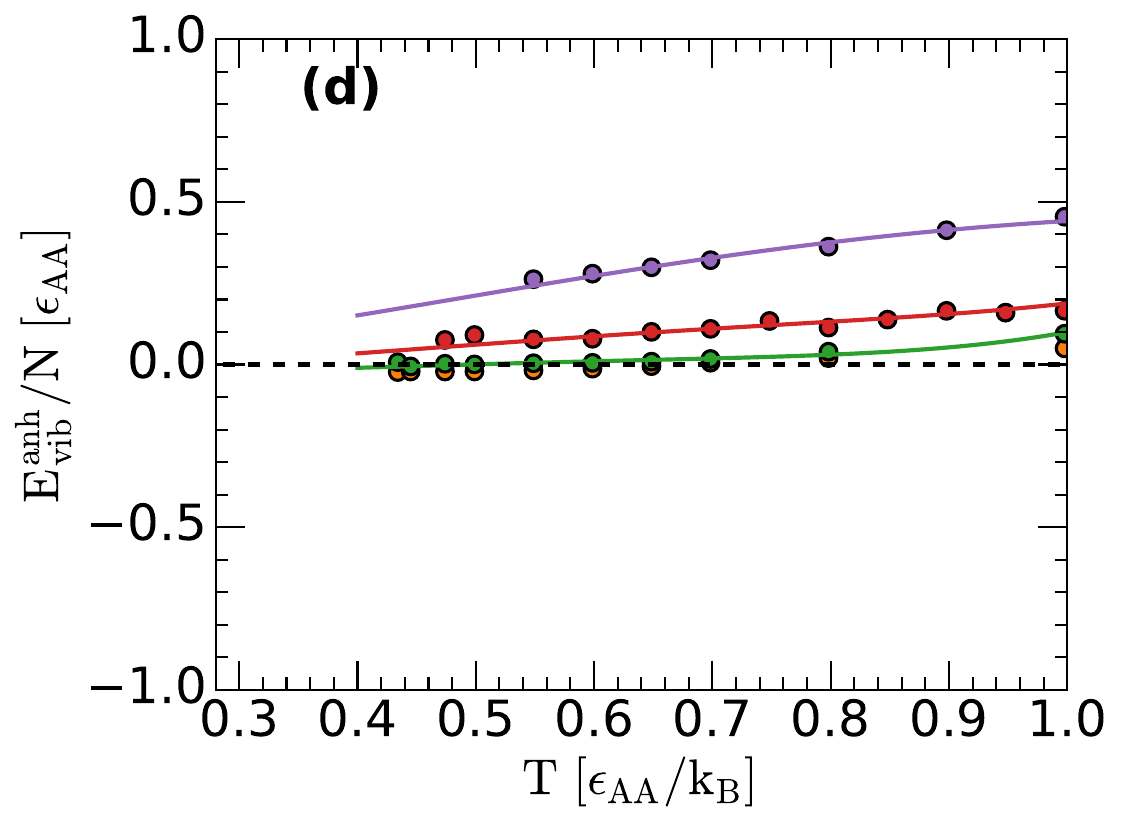}  
}
\caption{(a) Inherent structure energy $E_{IS}(T)$ and (b) vibrational energy $E_{vib}(T)$ as a function of temperature for the LJBMs studied. Results are obtained from RPMD simulations using a Planck's constant $h=0$ (i.e., classical MD simulations), and $h=h_a, h_b, h_c$ (circles).  Lines are the predictions of the PEL formalism for a Gaussian and harmonic PEL, $E_{IS}^{harm}(T)$ and $E_{vib}^{harm}(T)$, given in Eqs.~\ref{Eis_GAHA} and \ref{Evib_GAHA}. (c) Anharmonic contributions to the IS energy, $E_{IS}^{anh}(T)= E_{IS}(T) - E_{IS}^{harm}(T)$, obtained from (a) [circles]. Lines are the fit to the data points using Eq.~\ref{Eis_GAANH_expli} and \ref{Bj}. (d) Anharmonic contributions to the vibrational energy, $E_{vib}^{anh}(T)= E_{vib}(T) - E_{vib}^{harm}(T)$, obtained from (b) [circles]. Lines are the fit to the data points using Eq.~\ref{Evib_GAANH_expli} and \ref{Bj}. Anharmonicities are relevant for $h=h_b,h_c$ but not for $h=0,h_a$. Dashed-lines in (c)(d) correspond to zero energy.}
\label{fig:Eis_Evib_sim_HA}
\end{figure}

{\it PEL of the LJBMs}.  
The IS and vibrational energy of the studied LJBMs, $E_{IS}(T)$ and $E_{vib}(T)$, are 
included in Fig.~\ref{fig:Eis_Evib_sim_HA}(a)(b).  
The results from the RPMD simulations are indicated by circles; the lines are the prediction using
Eqs~\ref{Eis_GAHA} and \ref{Evib_GAHA}  for the case where the PEL of the classical/quantum LJBMs are assumed to be Gaussian and {\it harmonic}.  As explained in Sec.~\ref{sec:hypotheses}, we consider that the distribution of IS in the CL-PEL 
and RP-PEL are identical, i.e., in both cases, $\Omega_{IS}$ is given by Eq.~\ref{GaussPEL} with the {\it same}
values of $\{ \alpha, E_0, \sigma^2 \}$.  It follows from Fig.~\ref{fig:Eis_Evib_sim_HA}(a)(b) that, 
when nuclear quantum effects are small ($h=0,h_a$), the Gaussian and harmonic approximation of the PEL are
fully consistent with the RPMD simulations of the LJBMs. However, as the quantum nature of the LJBM increases ($h=h_b,h_c$), 
deviations between the predictions of the PEL formalism (within the Gaussian and harmonic approximation of the PEL, Eqs~\ref{Eis_GAHA} and \ref{Evib_GAHA}) and RPMD simulations become evident in both $E_{IS}(T)$ and $E_{vib}(T)$. 
Accordingly, for $h=h_b,h_c$, anharmonic corrections are needed.

\begin{figure}[!htb]
    \centering{
        \includegraphics[width=0.9\textwidth]{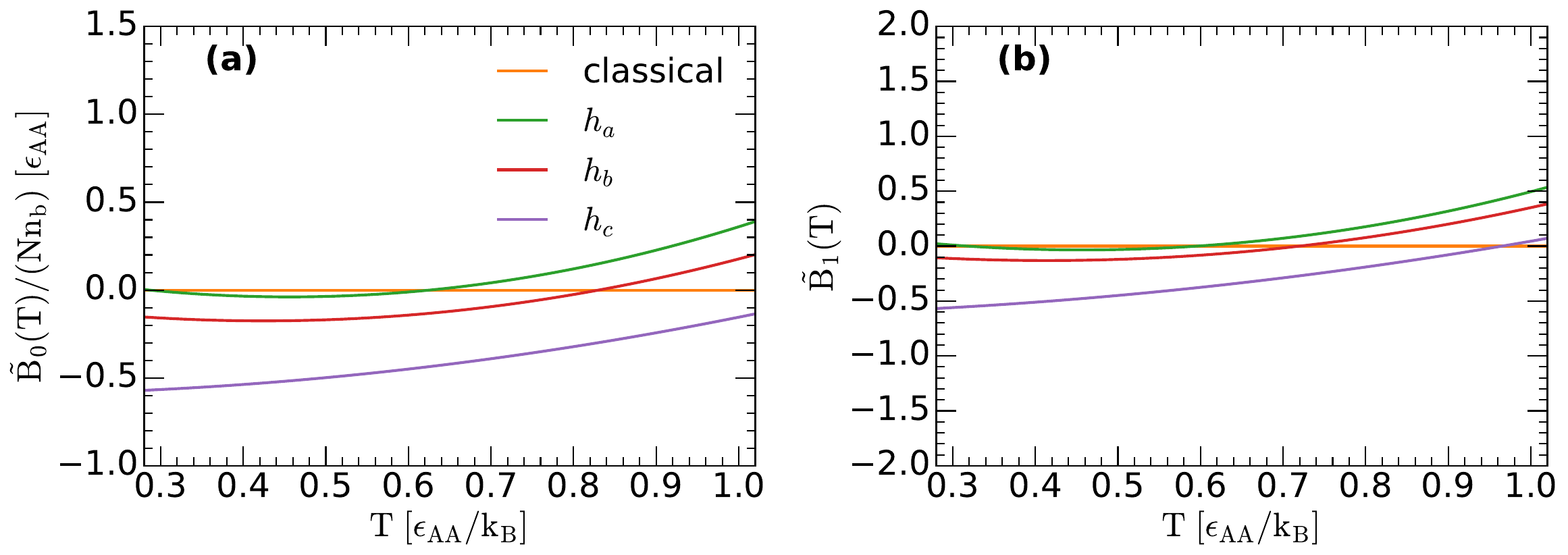}
    }
    \caption{Coefficient $\tilde{B}_0$ and $\tilde{B}_1$ defined in Eq.~\ref{Fanh_expli} and obtained from the fittings in Figs.~\ref{fig:Eis_Evib_sim_HA}(c)(d).}
    \label{fig:tilde_B0_B1}
    \end{figure}

The anharmonic corrections to $E_{IS}(T)$ and $E_{vib}(T)$ can be calculated numerically from Figs.~\ref{fig:Eis_Evib_sim_HA}(a)(b) and using the expressions 
 $E^{anh}_{IS}(T)=  E_{IS}(T) -  E^{harm}_{IS}(T)$ and $E^{anh}_{vib}(T)=  E_{vib}(T) -  E^{harm}_{vib}(T)$.
The so-obtained values of $E^{anh}_{IS}(T)$ and $E^{anh}_{vib}(T)$ are indicated by circles in 
Fig.~\ref{fig:Eis_Evib_sim_HA}(c)(d). Also included in Figs.~\ref{fig:Eis_Evib_sim_HA}(c)(d) are the fits to $E_{IS}^{anh}(T)$ and $E_{vib}^{anh}(T)$ using Eqs.~\ref{Eis_GAANH_expli} and \ref{Evib_GAANH_expli},  respectively, and  Eq.~\ref{Bj} (lines).
The fits to the data points in Figs.~\ref{fig:Eis_Evib_sim_HA}(c)(d) are excellent indicating that, for the LJBMs studied, the anharmonic contributions to the vibrational free energy can indeed be modeled using Eq.~\ref{Fanh_expli} with the coefficients ${\tilde B}_j$ given by Eq.~\ref{Bj}. The coefficients $B_0$ and $B_1$ are shown in Fig.~\ref{fig:tilde_B0_B1}.

For comparison, we include in Fig.~\ref{fig:Eis} the values of $E_{IS}(T)$ and $E_{vib}(T)$ obtained numerically (taken from Fig.~\ref{fig:Eis_Evib_sim_HA}(a)(b); circles) together with the corresponding predictions of the PEL formalism {\it including} the anharmonic corrections; see Eqs.~\ref{EisFinal} and \ref{EvibFinal}. 
The agreement between the predictions of the  PEL formalism (with the Gaussian approximation and including anharmonicities) and RPMD simulations is excellent.
Accordingly, for the case of the LJBMs studied, one can model the anharmonicities of the PEL as indicated by Eqs.~\ref{Fanh_expli} and \ref{Bj}.  Importantly, Fig.~\ref{fig:Eis} also supports strongly that hypothesis (B) indeed holds for the case of the LJBMs studied.

{\it Configurational Entropy.}
Next, we study the configurational entropy of the quantum LJBMs.  We also test whether the hypotheses of this work are self-consistent within 
the PEL formalism.  Specifically, our results are based on the hypotheses that (i) $\Omega_{IS}(e_{IS})$ is identical for the classical and quantum liquids (given by the Gaussian approximation, Eq.~\ref{GaussPEL}, with identical PEL variables $\{ \alpha, E_0, \sigma^2 \}$), and that (ii) the PEL anharmonicities can be modeled using Eq.~\ref{Fanh_expli}.

\begin{figure}[!htb]
\centering{
    \includegraphics[width=0.48\linewidth]{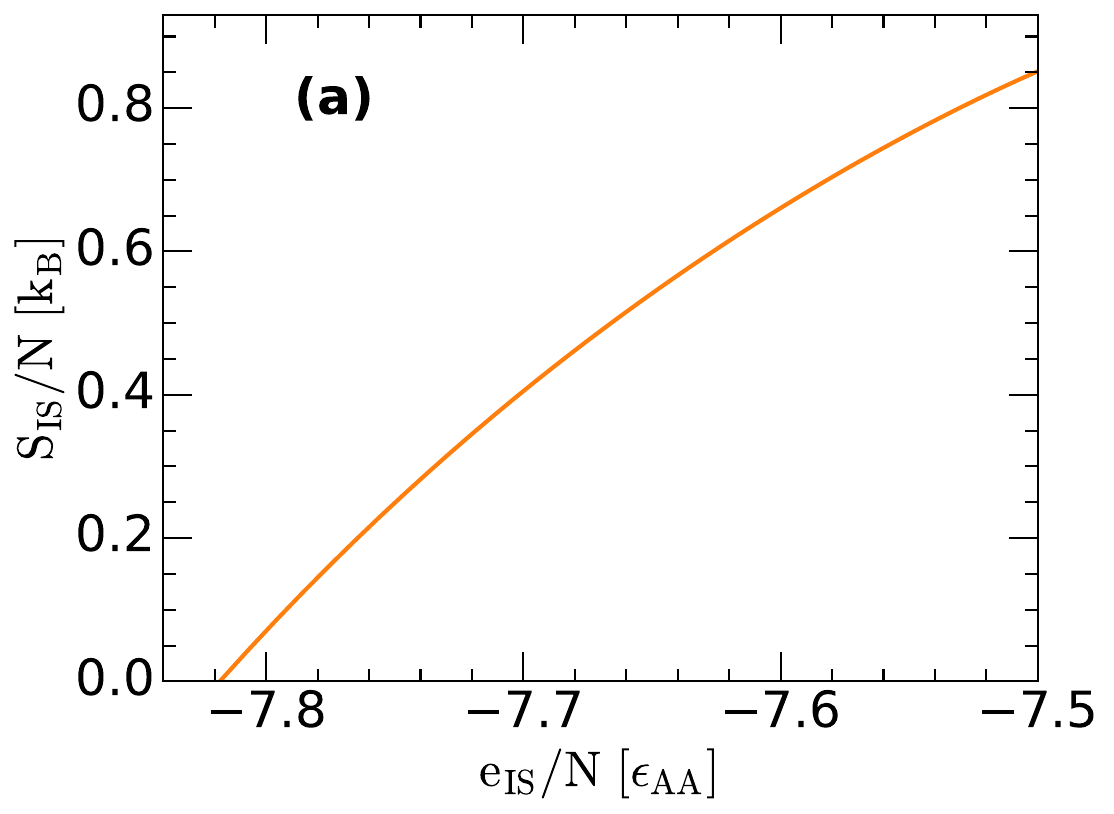}  
    \includegraphics[width=0.48\linewidth]{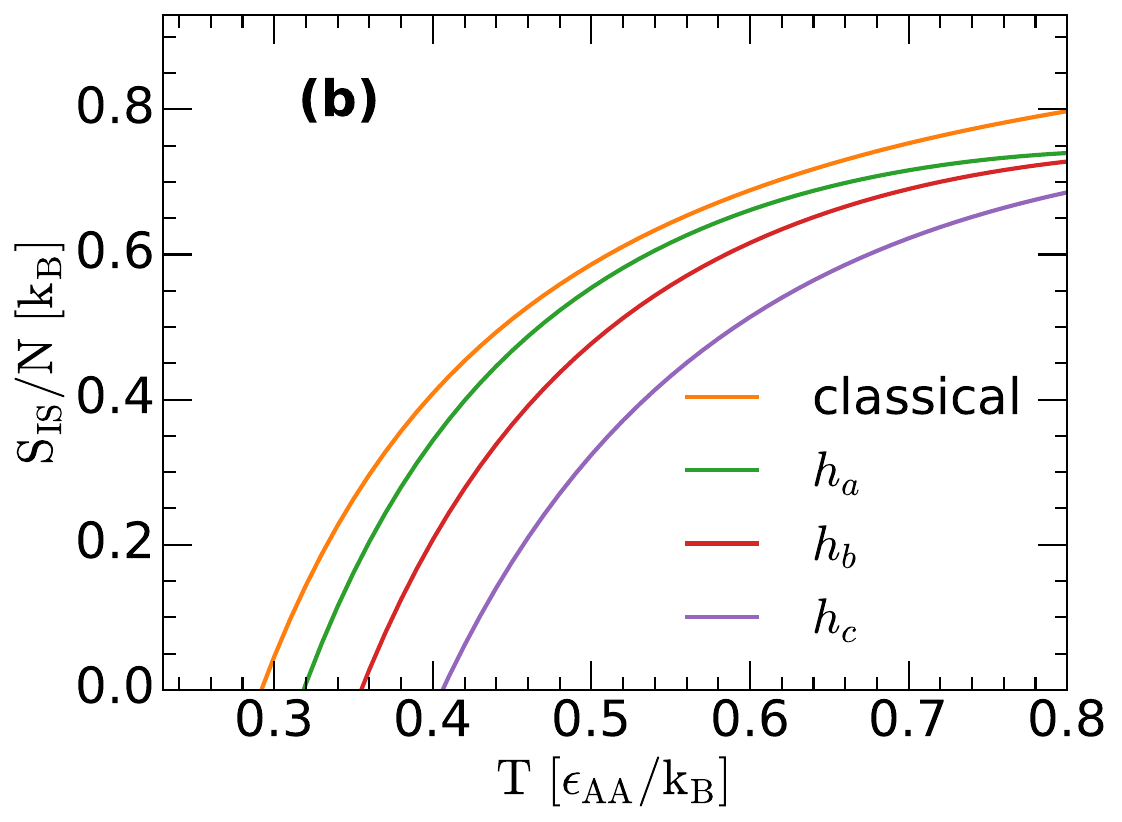}
}  
    \caption{(a) Configurational entropy as a function of the IS energy, $S_{IS}(e_{IS})$, for the classical and quantum LJBMs studied [Eq.~\ref{SconfGauss}]; $S_{IS}(e_{IS})$ is independent of the quantum character of the LJBM considered (see text). 
    (b) Configurational entropy as a function of temperature, $S_{IS}(T)=S_{IS}(E_{IS}(T))$. The $S_{IS}(T)$ for the LJBMs studied are obtained from (a) [Eq.~\ref{SconfGauss}] and using Eq.~\ref{EisFinal}.
}    
\label{fig:Sconf_T}
\end{figure}

Hypothesis (i) implies that for all the LJBMs studied, $S_{IS}(N,V,T,e_{IS})= S_{IS}(N,V,e_{IS})$, independent of $h$, with $S_{IS}(N,V,e_{IS})$ given by Eq.~\ref{SconfGauss}.  Again, this implies that,
{\it at a given depth $e_{IS}$ of the corresponding PEL}, the classical and quantum liquids have access to the same number of IS  [Fig.~\ref{fig:Sconf_T}(a) shows the $S_{IS}(N, V, e_{IS})$ of the LJBMs as a function of $e_{IS}$].
However, the number of IS accessible to the LJBMs {\it at a given temperature} does depend on the quantumness of the LJBM considered. To show this, we include in Fig.~\ref{fig:Sconf_T}(b),  $S_{IS}$ as a function of $T$, after substituting $E_{IS}(T)$ using Eq.~\ref{EisFinal}. Since $E_{IS}(T)$ varies with $h$ (Fig.~\ref{fig:Eis}), the temperature-dependence of $S_{IS}$ is different for the LJBMs studied. Our RPMD simulations show that $S_{IS}$ shifts towards higher temperatures as $h$ increases. In particular, the Kauzmann temperature $T_{K}$, defined as the temperature at which $S_{IS}=0$, increases considerably as the liquid becomes more quantum. Specifically, $T_K=0.291, 0.319, 0.355, 0.406$ for $h=0,h_a,h_b,h_c$, respectively. Overall, our results imply that, \textit{at a given temperature}, the LJBMs explore IS located at different depths of the corresponding PEL (Fig.~\ref{fig:Eis}) and hence, they have access to a different number of IS in the corresponding PEL (Fig.~\ref{fig:Sconf_T}(b)).

\begin{figure}[!htb]
\centering{
      
    \includegraphics[width=0.45\textwidth]{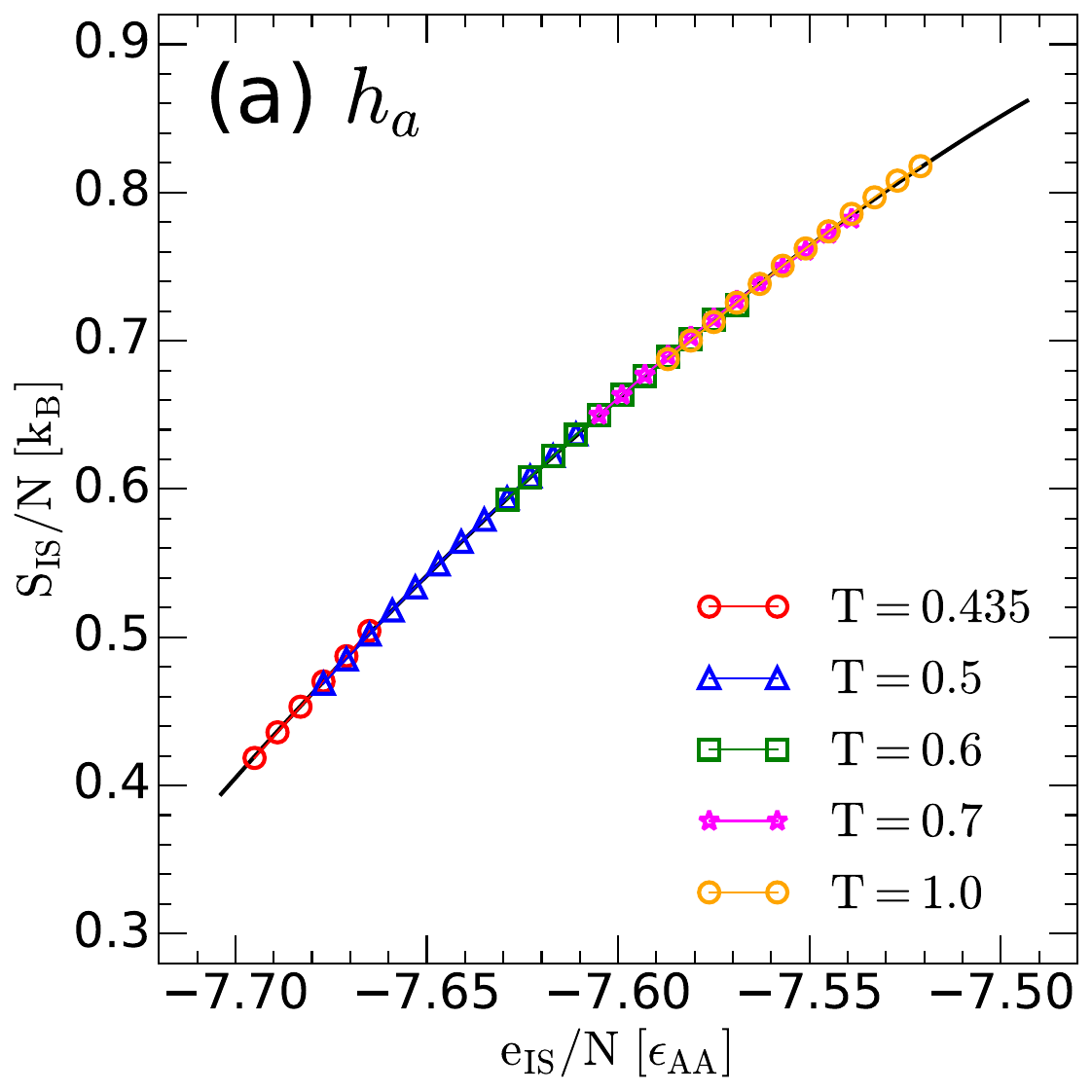}       
    \hfill
    \includegraphics[width=0.45\textwidth]{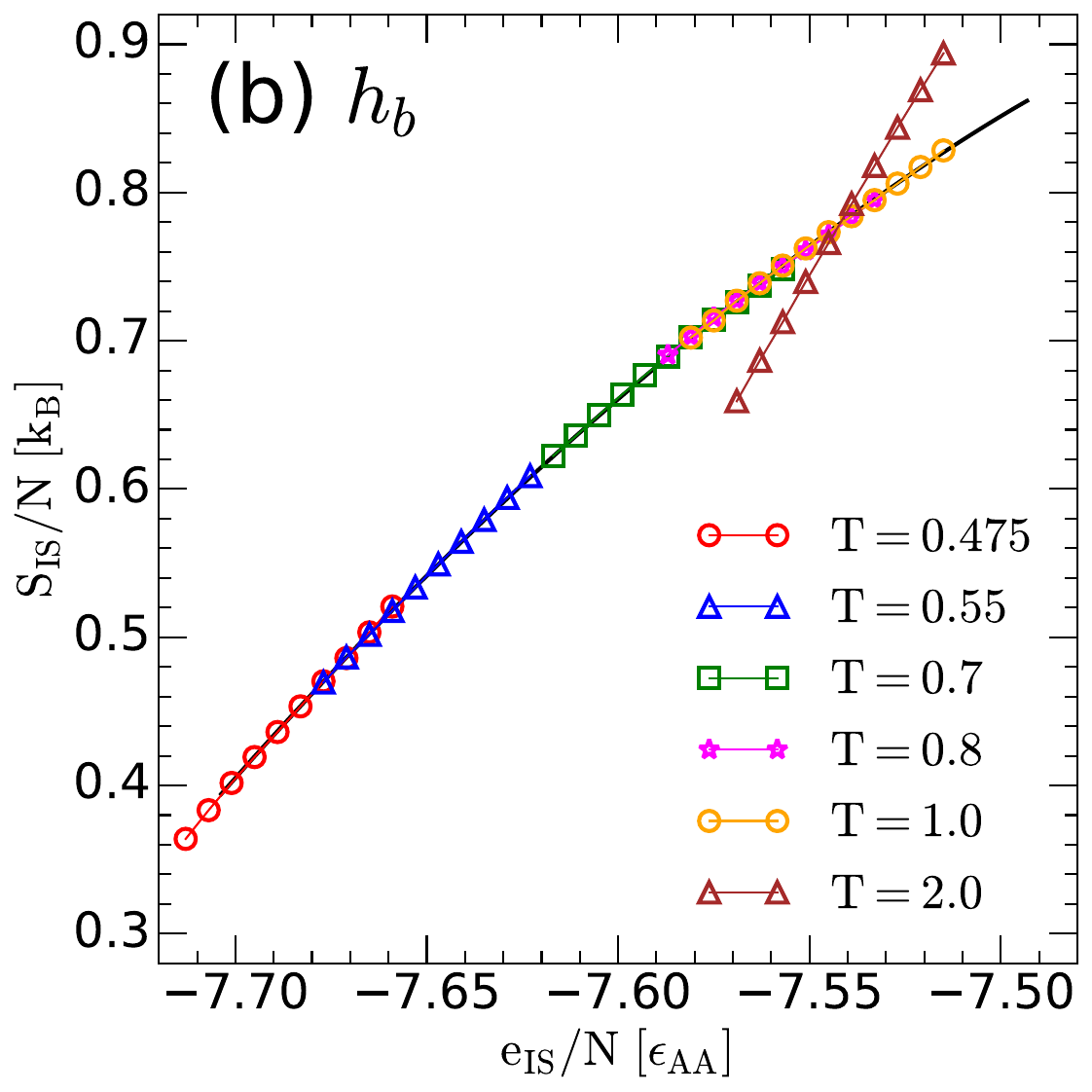}
    \hfill
    \includegraphics[width=0.45\textwidth]{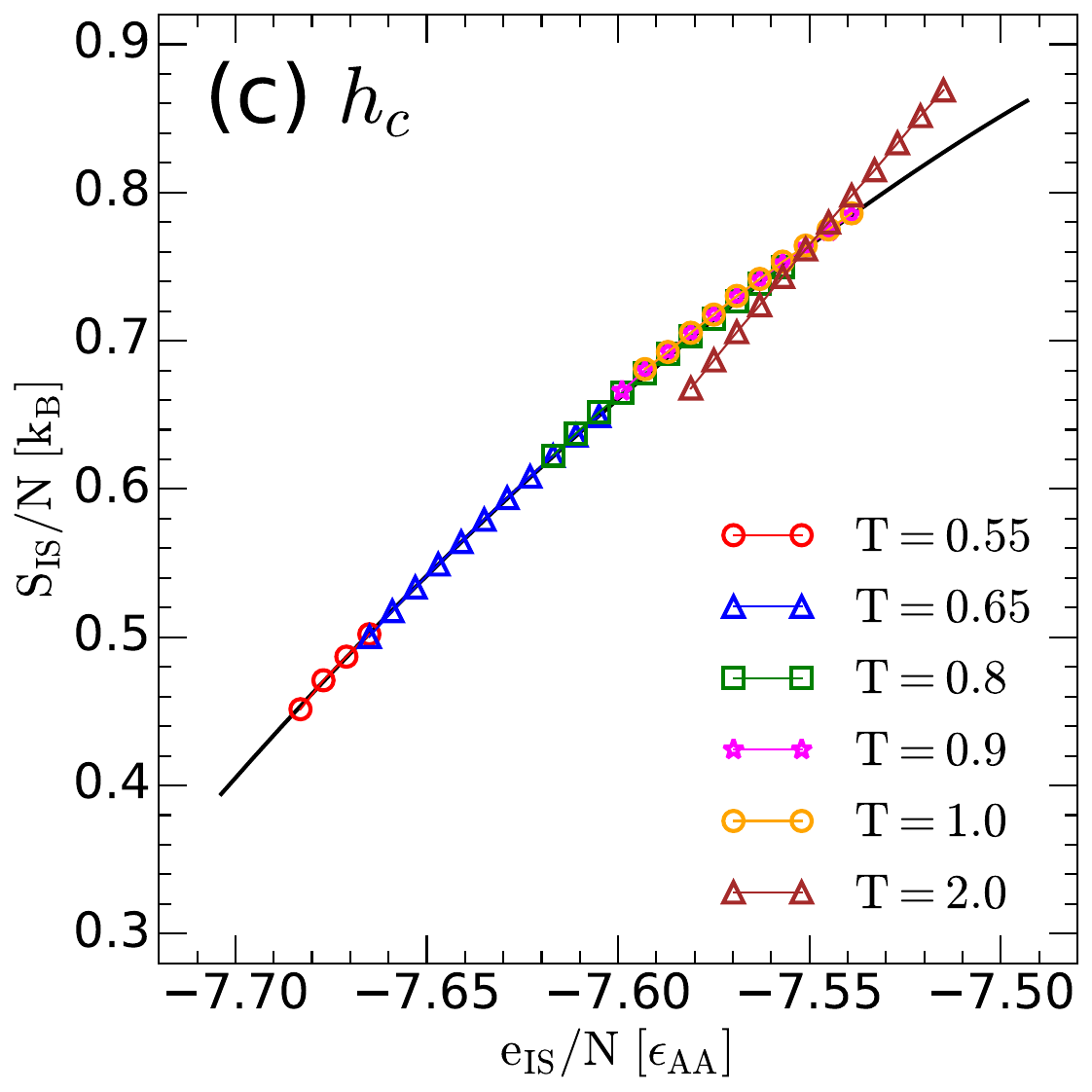}
}    
    \caption{Validation of the configurational entropy for the LJBMs. (a) Configurational entropy of the LJBMs as a function of the IS energy $e_{IS}$, $S_{IS}(e_{IS})$ [solid black line; taken from Fig.~\ref{fig:Sconf_T}(a)]. The symbols correspond to the expression on the right-hand-side of Eq.~\ref{SconfTEST1} for temperatures $T=0.435$ (red circles), $0.5$ (blue up-triangles), $0.6$ (green squares), $0.7$ (magenta stars) and $1.0$ (orange circles). Results at the different temperatures are obtained from RPMD simulations with $h=h_a$ [datasets are shifted by a constant $c(T)$; see text]. (b) Same as (a) for $h=h_b$, and for temperatures $T=0.475$ (red circles), $0.55$ (blue up-triangles), $0.7$ (green squares), $0.8$ (magenta stars), $1.0$ (orange circles), and  $2.0$ (maroon up-triangles). (c) Same as (a) for $h=h_c$, and for temperatures $T=0.55$ (red circles), $0.65$ (blue up-triangles), $0.8$ (green squares), $0.9$ (magenta stars), $1.0$ (orange circles), and  $2.0$ (maroon up-triangles). In all cases, the symbols fully overlap with the black line at low temperatures ($T\leq 1.0$), implying that Eq.~\ref{SconfTEST1} holds. At higher temperatures, deviations from Eq.~\ref{SconfTEST1} become evident [see, maroon up-triangles in (b) and (c) for the case $T=2.0$]}
    \label{fig:validate-Sconf}
\end{figure}

To confirm that the calculated configurational entropy of the LJBMs (Fig.~\ref{fig:Sconf_T}) 
are physically sound (i.e., self-consistent with the PEL formalism), we test that our results are consistent with Eq.~\ref{logPeis}. A similar self-consistency test was 
implemented for the case of the classical LJBM in Ref.~\cite{sciortinoInherentStructureEntropy1999,sciortinoThermodynamicsSupercooledLiquids2000}.  
Eq.~\ref{logPeis} imposes a strict relationship between the  $S_{IS}(N,V,e_{IS})$ 
and the probability distribution $\mathrm{P}(e_{IS}, T)$.  Specifically, at fixed $(N,V)$, the PEL formalism (Eq.~\ref{logPeis}) requires that
\begin{equation}    
\begin{aligned}
    S_{IS}(T,e_{IS})/k_B &= \ln \left( \mathrm{P}(T,e_{IS}) \right) + 3 N n_b \ln(\beta \hbar \omega_0) \\
    & + {\cal S}(T,e_{IS}) + \beta e_{IS} + \beta F^{anh}_{vib}(T, e_{IS}) + c(T)
    \label{SconfTEST1}
\end{aligned}
\end{equation}
where $c(T)$ is a quantity independent of $e_{IS}$.  
Accordingly, independently of the temperature considered, the right-hand-side of Eq.~\ref{SconfTEST1} should overlap with $S_{IS}(N,V,e_{IS})$ (up to some function $c(T)$). We note that $\mathrm{P}(T,e_{IS})$ is calculated numerically from the RPMD simulations and hence, Eq.~\ref{SconfTEST1}
provides a strong test to the validity of the $S_{IS}(N,V,e_{IS})$ reported in this work as well as to the underlying hypotheses (i) and (ii) stated above.
Fig.~\ref{fig:validate-Sconf} shows the $S_{IS}(N,V,e_{IS})$ of the LJBMs for $h=h_a,h_b,h_c$ 
[black line; taken
from Fig.~\ref{fig:Sconf_T}(a)] together with the corresponding fit given by the right-hand-side of Eq.~\ref{SconfTEST1} (the value of $c(T)$ is adjusted for a maximum overlap between both sides of Eq.~\ref{SconfTEST1}) \cite{sciortinoThermodynamicsSupercooledLiquids2000}.  It follows from Figure~\ref{fig:validate-Sconf} that Eq.~\ref{SconfTEST1} holds remarkably well up to approximately $T=1.0$, which is a rather high temperature for the LJBMs studied.

{\it Adam-Gibbs Relation}
The Adam-Gibbs relation relates the diffusion coefficient $D(T)$ of a liquid with its configurational entropy as follows ~\cite{adamTemperatureDependenceCooperative1965},
\begin{eqnarray}
D(T)= D_0 \exp \left( -A/T S_{IS} \right)
\label{AGeq}
\end{eqnarray}
where $D_0$ and $A$ are constants.  Eq.~\ref{AGeq} implies that the diffusivity at a a given temperature is controlled by the corresponding number of IS available to the system in the PEL, i.e., the topography of the PEL controls the dynamics of the liquid of interest.   Eq.~\ref{AGeq} has been validated in computational studies of diverse classical liquids, including silica~\cite{saika-voivodFreeEnergyConfigurational2004} and water~\cite{handleAdamGibbsRelation2018}.
Here, we show that the Eq.~\ref{AGeq} also holds for the LJBMs studied, independently of the quantum character ($h$) of the mixtures.  

The diffusion coefficient is calculated from the mean-square displacement (MSD) of the ring-polymer centroids as a function of time. Specifically, for an atomistic liquid, MSD$(t) \approx 6 D t$ at long-times~\cite{allenComputerSimulationLiquids2017}.  Fig.~\ref{fig:Adam-Gibbs} shows the so-obtained values of $D(T)$ as a function $1/TS_{IS}(T)$ for
all the LJBMs studied [circles; $S_{IS}$ is taken from Fig.~\ref{fig:Sconf_T}(b)]. The solid lines are the linear fits to the data points using Eq.~\ref{AGeq}.  The agreement between the RPMD simulations and the Adam-Gibbs equation is remarkable good, extending for approximately four orders of magnitude in $D$.

\section{Summary and Discussion}
\label{SecSummary}

One of the main goals of this work is to show that, from a thermodynamic/statistical mechanics point of view, the PEL formalism provides a unifying description of classical and quantum liquids over a wide range of temperatures down to the glass state.
In the classical case, the PEL of a liquid is given by the potential energy function of the system as a 
function of the atoms coordinates, $U(\mathbf{r}_1,\mathbf{r}_2,...,\mathbf{r}_N)$,
and the system is represented by a point moving on such a PEL (CL-PEL).  In the quantum case, 
one can associate the PEL of the underlying ring-polymer system (RP-PEL) to the quantum liquid 
(for a fix number of beads per ring-polymer, $n_b$). As discussed in this work, 
the PEL associated to the quantum liquid is defined by Eq.~\ref{Urp}, and the
quantum liquid, again, can be represented by a point moving on the so-defined RP-PEL. 
Therefore, independently of whether the system obeys classical or quantum mechanics, 
the {\it liquid} is characterized by a point that describes a trajectory on the corresponding CL-PEL/RP-PEL  
as the system visits different basins over time. 
In the {\it glass} state, the representative point of the system would be 
limited to vibrational motion within a single basin of the CL-PEL/RP-PEL as molecules/atoms are unable to diffuse.

\begin{figure}[!htb]
\centering{
\includegraphics[width=0.85\textwidth]{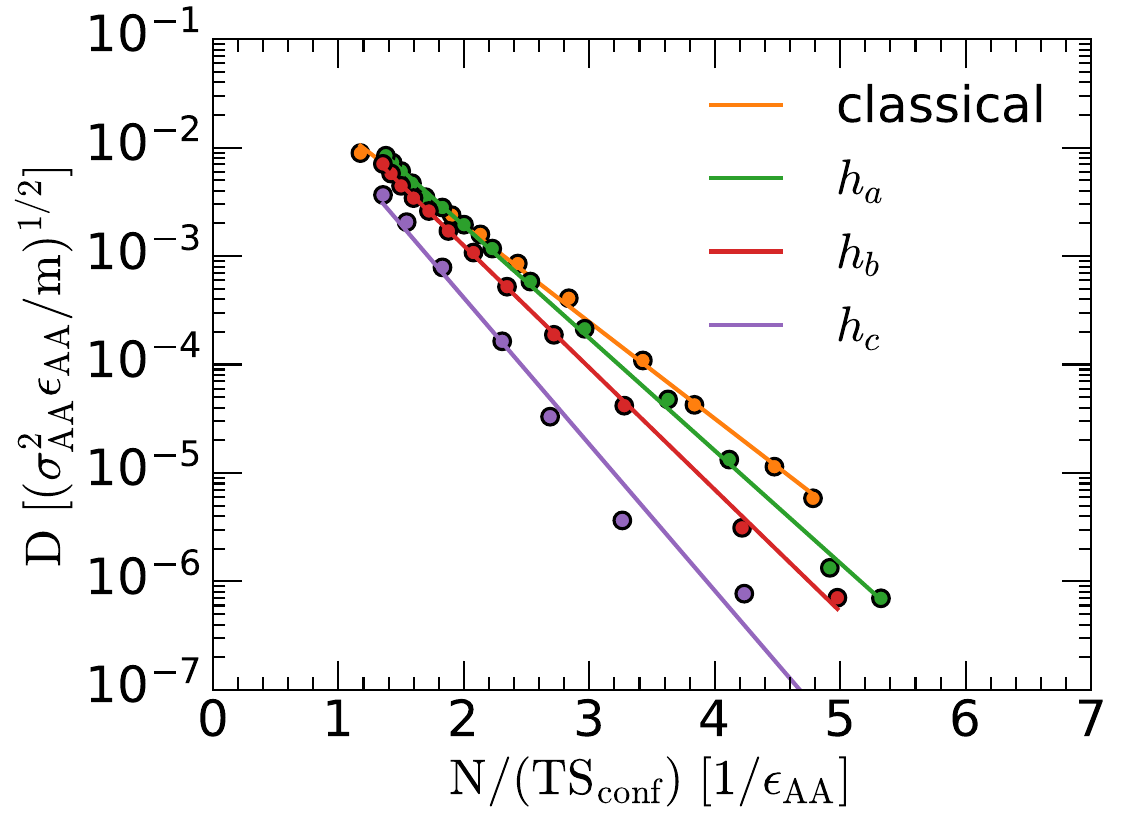}
}  
\caption{Diffusion coefficient of the LJBMs studied obtained from RPMD simulations using
$h=0$ (classical MD), and $h=h_a$, $h_b$, $h_c$  (circles).  Lines are the Adam-Gibbs prediction
(Eq.~\ref{AGeq}) with the configurational entropies $S_{IS}$ given in Fig.~\ref{fig:Sconf_T}(b).  The Adam-Gibbs relation holds for all the classical and quantum LJBMs studied for over more than four decades in $D$.}
\label{fig:Adam-Gibbs}
\end{figure}

Building upon the similarities between classical and quantum liquids in the PEL
formalism, we show that the RP-PEL can be used to define a
configurational entropy that can be associated to the quantum liquid in a precise manner. 
As for the classical case, $S_{IS}(N, V, T)$ is a measure of the number of IS available to the ring-polymer system
associated to the quantum liquid (Eq.~\ref{SconfGauss}). Importantly, using the so-defined configurational entropy,
it is shown that one can also extend the Adam-Gibbs relation to the quantum case. This implies
 that the dynamics of the quantum liquid is controlled by the
number of IS available to the ring-polymer system at a given $(N,V,T)$ [Fig.~\ref{fig:Adam-Gibbs}]. It follows that
the configurational entropy associated to the quantum liquid has a physical relevance from the dynamical
point of view.

The configurational entropy defined in this work is based on the hypothesis that (A) the distribution of IS, $\Omega_{IS}$, at a given working conditions $(N,V,T)$, is identical for the classical and quantum liquid.    
Within the Gaussian approximation of the PEL (Eq.~\ref{GaussPEL}), which is valid for the studied LJBMs (Fig.~\ref{fig:Eis}), this also 
implies that the PEL quantities $\{ \alpha, E_0, \sigma^2 \}$ do not depend on the nature (quantum/classical) of the liquid studied,
and that they depend only on $V$ (but not on $T$).  Accordingly, 
 the CL-PEL and RP-PEL have the same number of IS at a given IS energy/PEL depth. Hence, from a practical point of view,
the PEL quantities $\{ \alpha, E_0, \sigma^2 \}$ can be calculated from classical MD simulations (as opposed to the more 
computationally expensive path-integral computer simulations).
We stress that, as shown in Fig.~\ref{fig:validate-Sconf}, the so-obtained $S_{IS}(N,V,T)$ is validated by testing that
 Eq.~\ref{logPeis} is in agreement with the RPMD simulations of the LJBMs.   It follows that hypothesis (A) is self-consistent within 
the PEL formalism.  As explained in this work, hypothesis (A) should hold as
far as the ring-polymers associated to the quantum liquid collapse at the sampled IS.  
This seems to be rather general for systems where the NQE (atoms delocalization) are mild; indeed,
PIMD simulations of water~\cite{eltarebPotentialEnergyLandscape2024a}, monatomic 
quantum liquids~\cite{giovambattistaPotentialEnergyLandscape2020,zhouHarmonicGaussianApproximations2024,eltarebNuclearQuantumEffects2022}, and LJBM (this work) at low temperatures show no 
sign of delocalized ring-polymers at the corresponding IS. 

Our study emphasizes the role of anharmonic contributions in the PEL formalism.  For the case of the LJBMs with non-negligible quantumness ($h>h_a$), our RPMD simulations indicate that hypothesis (A) requires the inclusion of anharmonicities in the PEL-based description (Fig.~\ref{fig:Eis_Evib_sim_HA}). 
Accordingly, in this study, we also provide a general approach to include anharmonicities. In this regard, we note that Eqs.~\ref{EisAnh-GAANH} and \ref{EvibAnh-GAANH}  define unequivocally the {\it anharmonic} Helmholtz free energy since these are two independent equations that define the two partial derivatives of
$F_{vib}^{anh}(N,V,T,e_{IS})$ (at fixed $(N,V)$).  Our RPMD simulations based on the LJBMs show that while the PEL of the classical and weakly-quantum LJBMs are harmonic ($h \leq h_a$), anharmonicities 
become increasingly relevant with increasing $h$ ($h>h_a$).

Being able to extend the PEL formalism to low-temperature liquids/glasses that obey quantum mechanics provides a simple tool to explore the role of quantum mechanics on atomistic liquids composed of light-element as well as molecular liquids composed of small molecules that contain H atoms, such as water.  
In these cases, NQE can play a relevant role even at non-negligible temperatures (e.g., $T\approx80-270$~K)~\cite{ceriottiNuclearQuantumEffects2016}. 
Extending the concept of configurational entropy to quantum liquids is fundamental in providing a thermodynamic description of liquids that obey quantum mechanics.  For example, it may be possible to write an equation of state for quantum liquids using the PEL formalism by following a procedure similar to that used in the case of classical liquids~\cite{robertsEquationStateEnergy1999,handlePotentialEnergyLandscape2018}.

\section{Data Availability}
The authors confirm that the data supporting the findings of this study are available within the article and its supplementary material.

\section{Acknowledgments}
This work was supported by the NSF CREST Center for Interface Design and Engineered Assembly of Low Dimensional systems (IDEALS), NSF grant number HRD-1547380. This work was also supported by the Advanced Cyberinfrastructure Coordination Ecosystem: Services \& Support (ACCESS) program \cite{boernerACCESSAdvancingInnovation2023}, which is supported by the National Science Foundation Grant Nos. 2138259, 2138286, 2138307, 2137603, and 213829669.

\section{Additional Information}
The authors declare no conflict of interest.

\section{Supplementary Information}
In the Supplementary Information, we demonstrate the importance of the anharmonic contributions of the PEL by comparing the configurational entropy of LJBMs (Fig.~\ref{fig:Sconf_T}) using Eq.~\ref{logPeis} but omitting the anharmonic term $F_{vib}^{anh}$. Our result shows that the anharmonic term $F_{vib}^{anh}$ becomes significant for systems with stronger quantumness ($h_b,h_c$). The SI also includes results from RPMD simulations showing that Eq.~\ref{shape-b} holds for the LJBMs studied.


\section{Figures}

\newpage

\begin{figure}[!htb]
\centering{
    \includegraphics[width=0.45\textwidth]{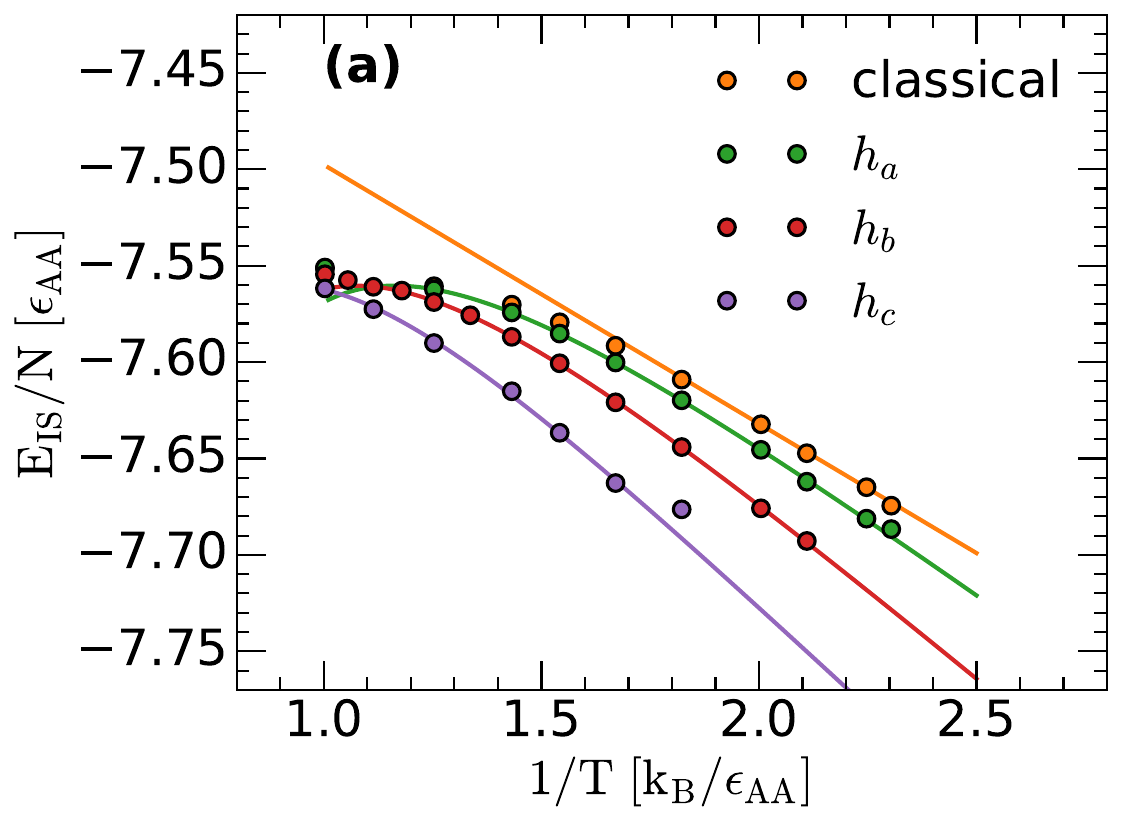}  
    \includegraphics[width=0.45\textwidth]{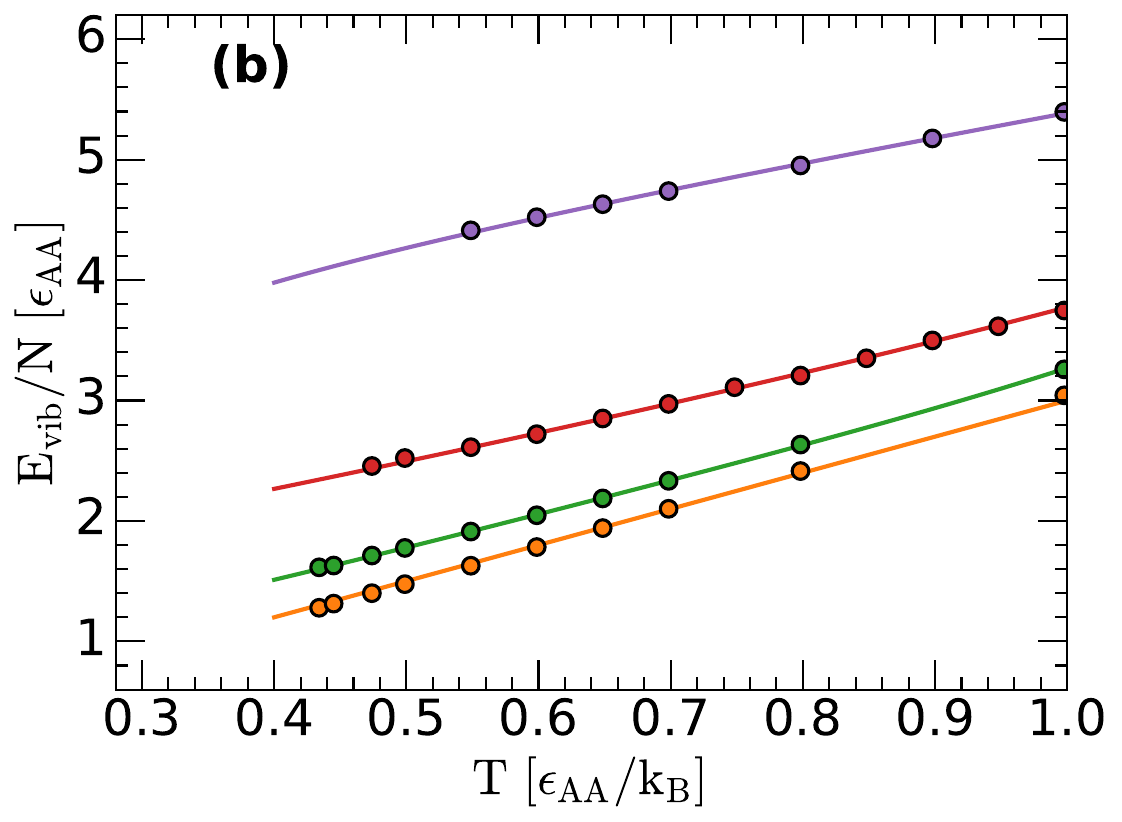}  
}
\caption{(a) Inherent structure energy $E_{IS}(T)$ and (b) vibrational energy $E_{vib}(T)$ as a function of temperature for the LJBMs studied. Results are obtained from RPMD simulations [circles; taken from Figs.~\ref{fig:Eis_Evib_sim_HA}(a) and \ref{fig:Eis_Evib_sim_HA}(b)]. Lines are the fit to the data using the predictions of the PEL formalism for a Gaussian and{\it anharmonic} PEL, using Eqs.~\ref{EisFinal} and \ref{EvibFinal} (see text).}
\label{fig:Eis}
\end{figure}

\newpage

\clearpage



\appendix

\section{Appendix: Inherent Structure ($E_{IS}$) and Vibrational Energy ($E_{vib}$) for a Gaussian Potential Energy Landscape}
\label{AppendixA}
\setcounter{equation}{0}
\renewcommand{\theequation}{A\arabic{equation}}

Here, we show how the formal expressions for $E_{IS}(N,V,T)$ and $E_{vib}(N,V,T)$ are obtained in the PEL formalism. The PEL is assumed to be Gaussian. We consider both cases where the harmonic approximation of the PEL holds (Eqs.~\ref{Eis_GAHA} and~\ref{Evib_GAHA}), and where anharmonicities are included (Eqs.~\ref{Eis_GAANH} and~\ref{Evib_GAANH}).

\textit{Harmonic PEL.}
The expression for \(E_{IS}\) (Eq.~\ref{Eis_GAHA}) follows from Eq.~\ref{EisDef}, using the expression for the configurational entropy of a Gaussian PEL (Eq.~\ref{SconfGauss}) and the vibrational Helmholtz free energy for a harmonic PEL (Eq.~\ref{FvibHA}). Specifically, substituting Eqs.~\ref{SconfGauss} and~\ref{FvibHA} into Eq.~\ref{EisDef}, one can show that,
\begin{equation}
1 - kT \left[ -\left( \frac{e_{IS}-E_0}{\sigma^2} \right) \right] + kT \left( \frac{\partial {\cal S}}{\partial e_{IS}} \right)_{N,V,T} = 0 \quad \text{at } e_{IS}=E_{IS}.
\label{eq:A}
\end{equation}
It follows that,
\begin{equation}
 E_{IS}(N,V,T) =  E_0(V) - \sigma^2(V) \left( \beta + b(N,V,T,E_{IS}) \right) 
\label{eq:B}
\end{equation}
where \( b(N,V,T,e_{IS}) \equiv \left( \frac{\partial {\cal S}(N,V,T,e_{IS})}{\partial e_{IS}} \right)_{N,V,T} \). Eq.~\ref{eq:B} defines \(E_{IS}\) implicitly. However, for most systems studied, including the LJBMs studied in this work, it is found that \( b = b(N,V,T) \) [see Eq.~\ref{shape-b}], and hence, Eq.~\ref{eq:B} can be used to define \(E_{IS}\) explicitly. To do so, one first needs to calculate \( {\cal S}(N,V,T,e_{IS}) \), from which \( b(N,V,T) \) is derived. Since the PEL variables \( E_0 \) and \( \sigma^2 \) are assumed to be \( T \)-independent (i.e., \( E_0 = E_0(V) \) and \( \sigma^2 = \sigma^2(V) \)), these variables can be obtained by plotting \( E_{IS}(T) \) as a function of \( \beta + b \), and fitting the data using Eq.~\ref{eq:B} [see, e.g., Fig.~\ref{fig:Eis}].

The expression for \( E_{vib} \) (Eq.~\ref{Evib_GAHA}) follows from the definition \( E_{vib} \equiv E - E_{IS} \) and the thermodynamic expression \( E = \left( \frac{\partial (\beta F)}{\partial \beta} \right)_{N,V} \). Since \( F = -kT \ln Q \), it follows from Eq.~\ref{Qfinal} that,
\begin{equation}
E(N,V,T) = \left(\frac{\partial}{\partial \beta} \left[ \beta E_{IS} - \frac{1}{k_B} S_{IS} + \beta F_{vib} \right]\right)_{N,V}
\label{eq:C}
\end{equation}
The first two terms of Eq.~\ref{eq:C} can be calculated using Eq.~\ref{SconfGauss} (Gaussian approximation),
\begin{equation}
\left( \frac{\partial}{\partial \beta} \left[ \beta E_{IS} - \frac{1}{k_B} S_{IS} \right] \right)_{N,V} = \left( \frac{\partial}{\partial \beta} \left[ \beta E_{IS} - \left( \alpha N - \frac{(E_{IS}-E_0)^2}{2 \sigma^2} \right)\right]\right)_{N,V}
\label{eq:D}
\end{equation}
Using Eq.~\ref{eq:B} in Eq.~\ref{eq:D}, one obtains the following expression,
\begin{equation}
\left(\frac{\partial}{\partial \beta} \left[ \beta E_{IS} - \frac{1}{k_B} S_{IS} \right]\right)_{N,V} = E_{IS} - N \left( \frac{\partial\alpha}{\partial \beta} \right)_{V} + \beta \left( \frac{E_0}{\partial \beta} \right)_{V} + \frac{1}{2} \left( \frac{\partial \sigma^2}{\partial \beta} \right)_{V}  (\beta^2 + b^2) + b \sigma^2 \left[ 1 + \left( \frac{\partial b}{\partial \beta} \right)_{N,V} \right]
\label{eq:E}
\end{equation}
The last term of Eq.~\ref{eq:C} can be calculated using Eq.~\ref{FvibHA} (harmonic approximation),
\begin{equation}
\left( \frac{\partial (\beta F_{\text{vib}})}{\partial \beta} \right)_{N,V} = 3N n_b k_B T + \left[ \frac{\partial {\cal S}(N,V,T,E_{IS})}{\partial \beta} \right]_{N,V}
\label{eq:F}
\end{equation}
Since \( {\cal S} = {\cal S}(N,V,T,E_{IS}) \), it follows that
\begin{equation}
\begin{aligned}
\left( \frac{\partial {\cal S}}{\partial \beta} \right)_{N,V} &= \left( \frac{\partial {\cal S}}{\partial 
\beta} \right)_{N,V,E_{IS}} + \left( \frac{\partial {\cal S}}{\partial E_{IS}} \right)_{N,V,\beta} \left( \frac{\partial E_{IS}}{\partial \beta} \right)_{N,V}\\
&=\left( \frac{\partial {\cal S}}{\partial 
\beta} \right)_{N,V,E_{IS}} + b \left( \frac{\partial E_{IS}}{\partial \beta} \right)_{N,V}
\label{eq:G}
\end{aligned}
\end{equation}
Using Eq.~\ref{eq:B} in Eq.~\ref{eq:G}, and then replacing the result into Eq.~\ref{eq:E}, one finds that
\begin{equation}
\left( \frac{ \partial(\beta F_{vib})}{\partial \beta} \right)_{N,V} = 3N n_b k_B T +  \left( \frac{\partial {\cal S}}{\partial \beta} \right)_{N,V,E_{IS}} + b \left[\left( \frac{\partial E_0}{\partial \beta} \right)_{V} - \sigma^2  \left(\frac{\partial }{\partial\beta}(\beta + b) \right)_{V} - \left( \frac{\partial \sigma^2}{\partial \beta} \right)_{V} (\beta + b) \right]
\label{eq:H}
\end{equation}
Using Eqs.~\ref{eq:E} and~\ref{eq:H} in Eq.~\ref{eq:C}, the following expression for the energy follows,
\begin{eqnarray}
E(N,V,T) = &E_{IS} + \left\{ 3N n_b k_B T +  \left( \frac{\partial {\cal S}}{\partial \beta} \right)_{N,V,E_{IS}} \right.\\ 
& \left. -N \left(\frac{\partial \alpha}{\partial \beta}\right)_{V}  +  (b+\beta)  \left(\frac{\partial E_0}{\partial \beta}\right)_{V} -  \frac{1}{2}(b+\beta)^2 \left(\frac{\partial \sigma^2}{\partial \beta}\right)_{V}   \right\}
\label{eq:I}
\end{eqnarray}
where the curly bracket is \( E_{vib} \equiv E - E_{IS} \). For the case where \( E_0 \), \( \sigma^2 \), and \( \alpha \) are \( T \)-independent, one finds that
\begin{equation}
E_{vib}(N,V,T) = 3N n_b k_B T + \left( \frac{\partial {\cal S}(N,V,T,E_{IS})}{\partial \beta} \right)_{N,V,E_{IS}} 
\label{eq:J}
\end{equation}

\textit{Anharmonic PEL.}
To obtain the expressions for \( E_{IS} \) and \( E_{vib} \), for a Gaussian and \textit{an}harmonic PEL, we follow the same procedure outlined above. To do so, we first note that, in the presence of anharmonicities, the vibrational Helmholtz free energy is given by Eq.~\ref{FvibAnh}, which we rewrite as follows,
\begin{equation}
F_{vib}(N,V,T,e_{IS}) \approx 3N n_b k_B T \ln(\beta \hbar \omega_0) + k_B T \left[ {\cal S}(N,V,T,e_{IS}) + \beta F_{vib}^{anh}(N,V,T) \right]
\label{eq:AA}
\end{equation}
The square bracket can be interpreted as an \textit{effective} shape function,
\begin{equation}
{\cal S}_{eff}(N,V,T,e_{IS}) \equiv {\cal S}(N,V,T,e_{IS}) + \beta F_{vib}^{anh}
\label{eq:XX}
\end{equation}
that takes into account the corrections due to the anharmonicities in the PEL basins, about the corresponding IS. This means that the vibrational Helmholtz free energy is given by Eq.~\ref{FvibHA}, substituting \( {\cal S}_{eff}(N,V,T,e_{IS}) \) for \( {\cal S}(N,V,T,e_{IS}) \),
\begin{equation}
F_{vib}(N,V,T) = 3N n_b kT \ln(\beta \hbar \omega_0) + k_B T {\cal S}_{eff}(N,V,T,e_{IS})
\label{eq:BB}
\end{equation}
Using Eq.~\ref{eq:BB}, we can obtain \( E_{IS} \) and \( E_{vib} \).

The expression for \( E_{IS} \) (Eq.~\ref{Eis_GAANH}) follows from Eq.~\ref{EisDef}, with the configurational entropy given by Eq.~\ref{SconfGauss} (Gaussian approximation), and the vibrational Helmholtz free energy given by Eq.~\ref{eq:BB}. Substituting Eqs.~\ref{SconfGauss} and~\ref{eq:BB} into Eq.~\ref{Qdef}, one can show that,
\begin{equation}
1 - kT \left[ -\frac{(e_{IS} - E_0)}{\sigma^2}\right] + kT \left( \frac{\partial {\cal S}_{eff}}{\partial e_{IS}} \right)_{N,V,T} = 0 \quad \text{at} \quad e_{IS} = E_{IS}
\label{eq:CC}
\end{equation}
Using Eq.~\ref{eq:XX} in Eq.~\ref{eq:CC}, it follows that,
\begin{equation}
E_{IS}(N,V,T) = E_0 - \sigma^2 
\left(b+\beta\right) - \sigma^2 \left( \frac{\partial \beta F_{vib}^{anh}}{\partial  e_{IS}} \right)_{N,V,T,e_{IS}=E_{IS}}
\label{eq:DD}
\end{equation}
which is identical to Eq.~\ref{Eis_GAANH}.

The expression for \( E_{vib} \) (Eq.~\ref{Evib_GAANH}) follows from the definition \( E_{vib} \equiv E - E_{IS} \) and the thermodynamic expression \( E = \left( \frac{\partial (\beta F)}{\partial \beta} \right)_{N,V} \). Since \( F = -kT \ln Q \), it follows from Eq.~\ref{Qfinal} that the total energy is given by Eq.~\ref{eq:C} even if anharmonicities are included. In the presence of anharmonicities, however, one must replace \( {\cal S}(N,V,T,e_{IS}) \) with \( {\cal S}_{eff}(N,V,T,e_{IS}) \) in the definition of \( F_{vib} \) [see Eqs.~\ref{FvibHA} and~\ref{eq:BB}]. The same mathematical steps followed in the case of harmonic PELs [Eqs.~\ref{eq:C}--\ref{eq:J}] can then be applied for anharmonic PELs. This implies that, in the presence of anharmonicities, Eq.~\ref{eq:J} holds after substituting  \( {\cal S}_{eff}(N,V,T,e_{IS}) \) for \( {\cal S}(N,V,T,e_{IS}) \). Therefore,

\begin{equation}
E_{vib} = 3N n_b kT + \left( \frac{\partial {\cal S}_{eff}(N,V,T,E_{IS})}{\partial \beta} \right)_{N,V,E_{IS}}
\label{eq:EE}
\end{equation}
Using Eq.~\ref{eq:XX} in Eq.~\ref{eq:EE}, one obtains the expression for \( E_{vib} \),
\begin{equation}
E_{vib}(N,V,T) = 3N n_b kT + \left( \frac{\partial {\cal S}(N,V,T)}{\partial \beta} \right)_{N,V,E_{IS}} + \left( \frac{\partial ( \beta F_{vib}^{anh}(N,V,T) )}{\partial \beta} \right)_{N,V,E_{IS}}
\label{eq:FF}
\end{equation}
which is identical to Eq.~\ref{Evib_GAANH}.

\section{Appendix: Relationship between the Temperature-dependence of the PEL of a quantum liquid and the {\it springs} potential energy of the corresponding ring-polymer system}

\label{AppendixB}
\setcounter{equation}{0}
\renewcommand{\theequation}{B\arabic{equation}}

Here we show that, within the harmonic approximation of the PEL, the potential energy of the ring-polymer's springs,
\[
E_{sp}(N,V,T)\equiv\left\langle\sum_{i=1}^N \sum_{k=1}^{n_b}  \frac{1}{2} k^{sp}_i (\mathbf{r}_i^{k+1} - \mathbf{r}_i^k)^2\right\rangle_{N,V,T}
\]
controls the temperature dependence of the PEL.  Specifically, we show that
\begin{equation}
\label{eqZZ}
\left( \frac{\partial {\cal S}(N,V,T,e_{IS})}{\partial \beta} \right)_{N,V,e_{IS}=E_{IS}} 
\;=\; 2\,E_{sp}(N,V,T)
\end{equation}

To do so, we introduce an imaginary ring-polymer (IRP) system as follows.  At a given temperature $T$, the ring-polymer system associated to the quantum liquid (briefly, “the” RP system) is characterized by a spring constant $k^{sp}(T)= m n_b/( \beta \hbar)^2$.  We define the IRP system so its spring constant $k_0$ is $T$-independent and given by $k_0 \equiv  k_{sp}(T)$.  Hence, while the PEL of the RP system (associated to the quantum liquid) varies with $T$, the PEL of the IRP does not.  The IRP system is a truly classical system in the sense that its PEL, and hence its Hamiltonian, do not vary with $T$.  Importantly, at the original temperature $T$, and only at this $T$, both IRP and the original RP share the same PEL, including the inherent structures available in the PEL.

Now, consider the evolution of the IRP (with constant spring constant $k_0$) and the RP associated to the quantum liquid [with spring constant $k^{sp}(T)$] upon heating/cooling.  The total energy of the original RP system is given by Eq.\ref{Evib_def} and \ref{Evib_GAHA} (harmonic approximation),
\begin{equation}
\label{eqAA}
E(N,V,T)= E_{IS}(N,V,T) \;+\; 3N\,n_b\,k\,T 
\;+\;\left( \frac{\partial {\cal S}(N,V,T,e_{IS})}{\partial \beta}\right)_{N,V,e_{IS}=E_{IS}}
\end{equation}
Instead, for the IRP ($T$-independent PEL), the total energy (harmonic approximation) is given by
\begin{equation}
\label{eqAB}
E^{IRP} \;=\; E_{IS}(N,V,T) \;+\; 3N\,n_b\,k\,T
\end{equation}
As shown in Ref.~\cite{tuckermanStatisticalMechanicsTheory2010}, 
\[
E(N,V,T)\;=\;E^{IRP}\;-\;2\,E^{sp}.
\]
Therefore, by substituting Eqs.~\ref{eqAA} and \ref{eqAB} into this expression, one obtains Eq.~\ref{eqZZ}.

\end{document}



\title{Configurational Entropy and Adam-Gibbs Relation for Quantum Liquids}


\author{Yang Zhou$^{1,2,*}$}
\author{Ali Eltareb$^{1,2,*}$}
\author{Gustavo E. Lopez$^{3,4,*}$}
\author{Nicolas Giovambattista$^{1,2,3}$}

\email{yzhou4@gradcenter.cuny.edu, gustavo.lopez1@lehman.cuny.edu, ngiovambattista@brooklyn.cuny.edu}
\affiliation{$^1$Department of Physics, Brooklyn College of the City University of New York, Brooklyn, New York 11210, United States \\
	$^2$Ph.D. Program in Physics, The Graduate Center of the City University of New York, New York, NY 10016, United States \\
	$^3$Ph.D. Program in Chemistry, The Graduate Center of the City University of New York, New York, NY 10016, United States\\
	$^4$Department of Chemistry, Lehman College of the City University of New York, Bronx, New York 10468, United States \\
}


\date{\today}


\maketitle


\setcounter{figure}{0}
\renewcommand{\thefigure}{S\arabic{figure}}

\setcounter{equation}{0}
\renewcommand{\theequation}{S\arabic{equation}}

\section{Linear Dependence of Shape Function on the Inherent Structure Energy}

\begin{figure}[!htb]
	\centering{
		\includegraphics[width=0.45\textwidth]{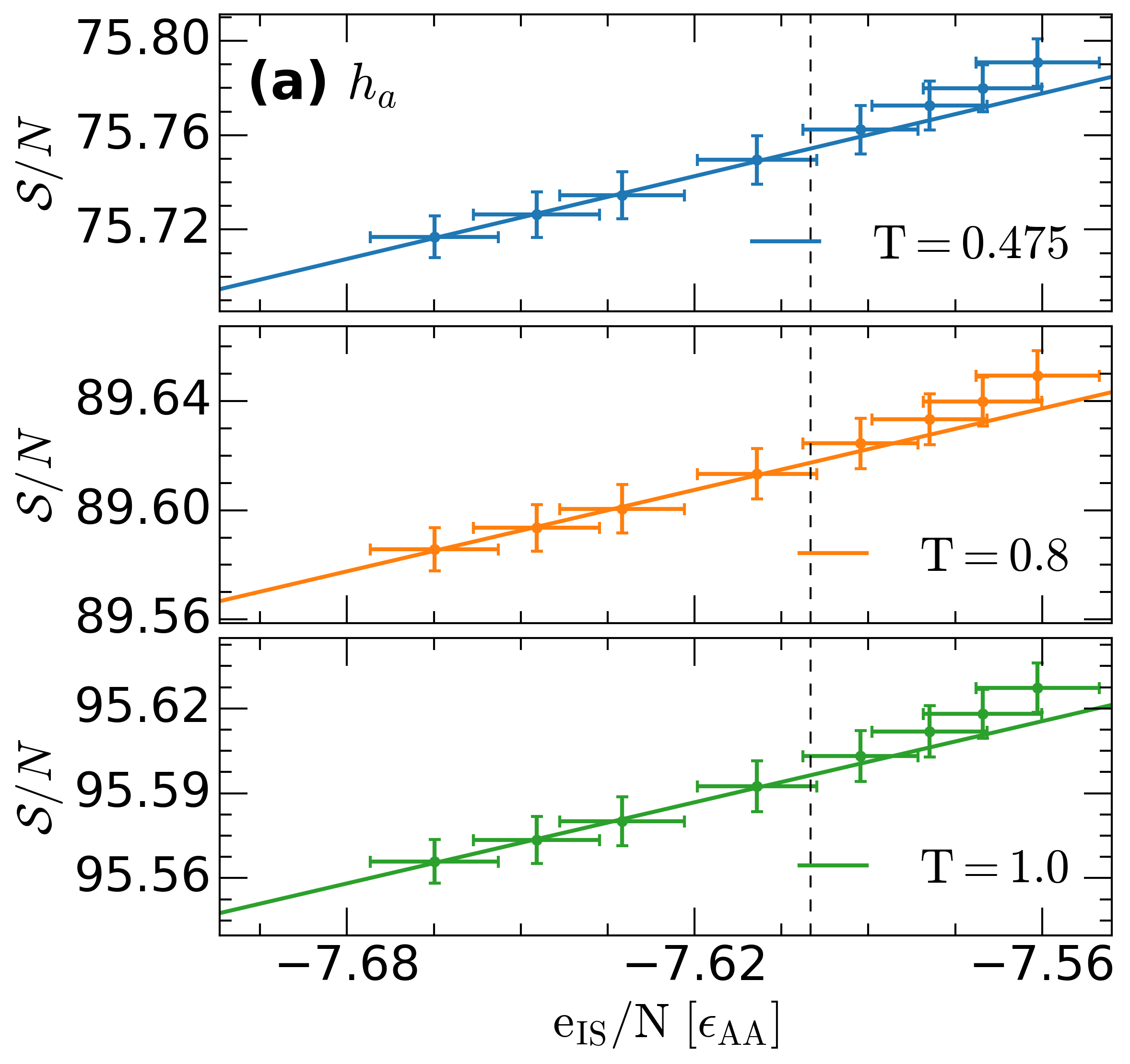}       
		\hfill
		\includegraphics[width=0.45\textwidth]{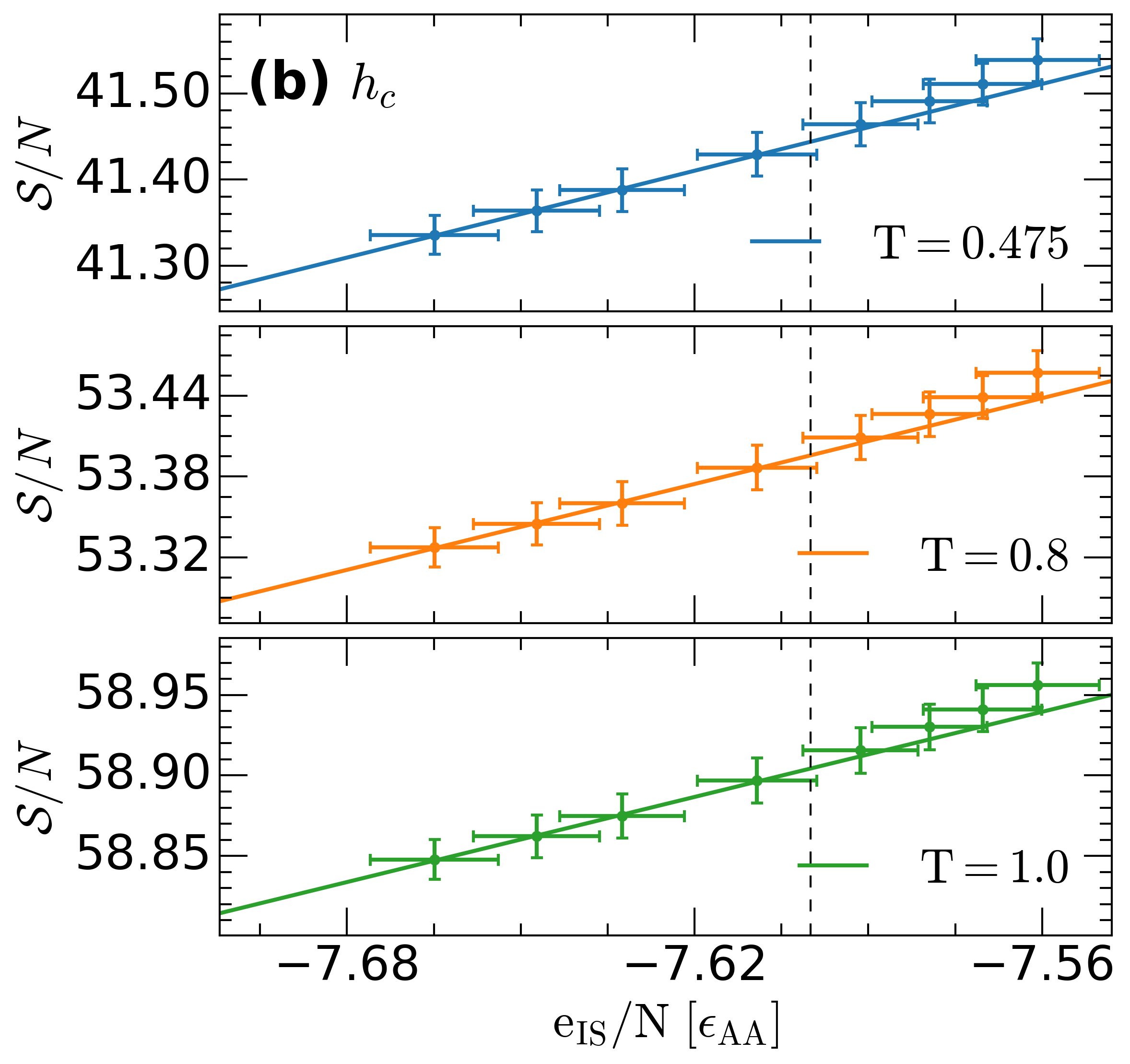}
	}    
	\caption{Shape function of the ring-polymer system as a function of the inherent structure energy for the LJBMs with Planck's constant (a) $h=h_a$ and (b) $h=h_c$. Each panel includes data at three different temperatures. The solid lines are linear fits to the data points with $e_{IS} \leq -7.6$ (dashed line). The solid lines are linear fits to the data points for $e_{IS} < -7.6$, where the Gaussian and harmonic approximation of the classical PEL holds.}
	\label{fig:linear-shapeFun}
\end{figure}

In this section, we show that the shape function $\mathcal{S}(N,V,T,e_{IS})$ for the ring polymers systems, defined in Eq.~12 in the main manuscript, is a linear function of the inherent structure energy $e_{IS}$ for all values of the quantumness $h$ and for all temperatures $T$ studied. In Fig.~\ref{fig:linear-shapeFun}, we plot the shape function of the ring-polymers system as a function of the inherent structure energy for the LJBMs with Planck's constant $h_a$ and $h_c$, at temperatures $T=0.475,0.8, 1.0$.  The shape function at each point is computed from the IS normal mode frequencies of the classical system, using the method described in Ref.~\cite{eltarebPotentialEnergyLandscape2024a}. Solid lines are linear fits to the data points with $e_{IS} \leq -7.6$ (dashed line), above which the Gaussian and harmonic approximation of the PEL of the classical system is no longer valid; see also Fig.~2 of the main manuscript. From the linear fits in Fig.~S1, we extract the fitting parameters $a(T)$ and $b(T)$ defined in Eq.~12 of the main manuscript.

\section{The Importance of Anharmonic Contributions to the Configurational Entropy}

The configurational entropy, $S_{IS}(N,V,e_{IS})$, of the classical and quantum LJBMs is given 
by Eq.~10 (Gaussian approximation) and is shown in Fig.~5(a).  In the main manuscript, this expression for $S_{IS}(N,V,e_{IS})$ (Eq.~10) is validated using Eq.~34 (or, equivalently, Eq.~33) including the anharmonic contributions to the Helmholtz free energy, $F_{vib}^{anh}(N,V,T)$. The agreement between Eq.~34 and Eq.~10 is remarkable (Fig.~6).

\begin{figure}[!htb]
	\centering{
  
		\includegraphics[width=0.45\textwidth]{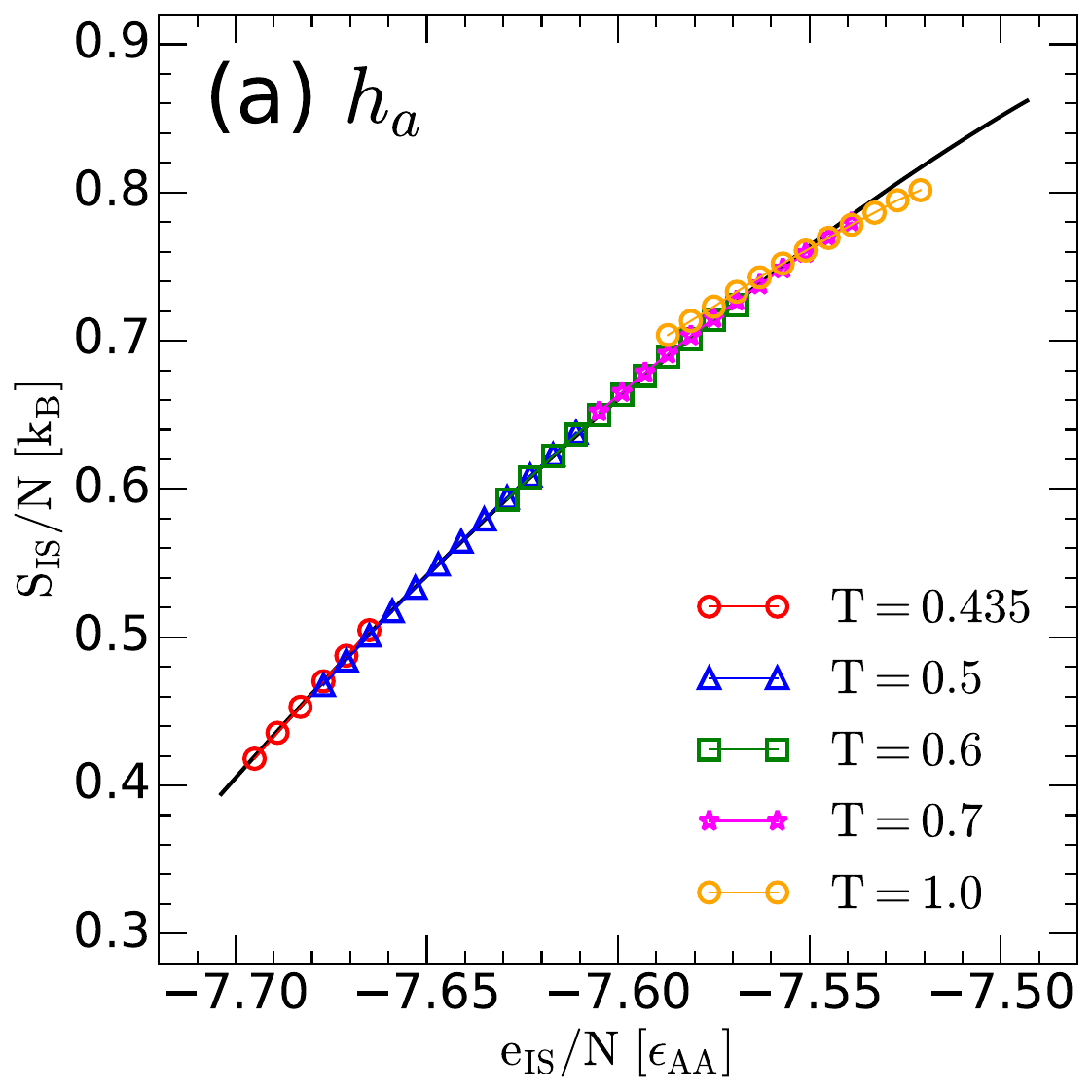}       
		\hfill
		\includegraphics[width=0.45\textwidth]{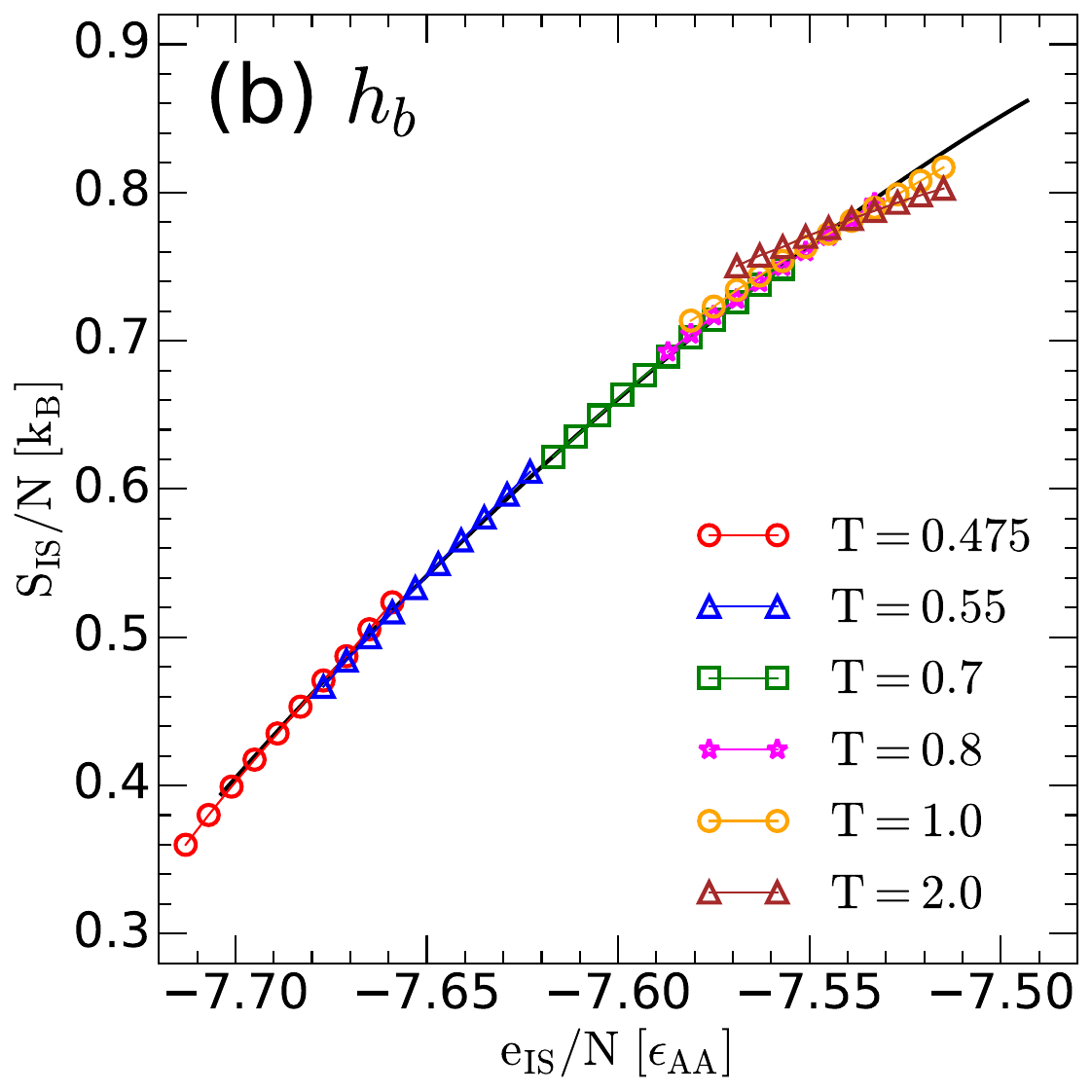}
		\hfill
		\includegraphics[width=0.45\textwidth]{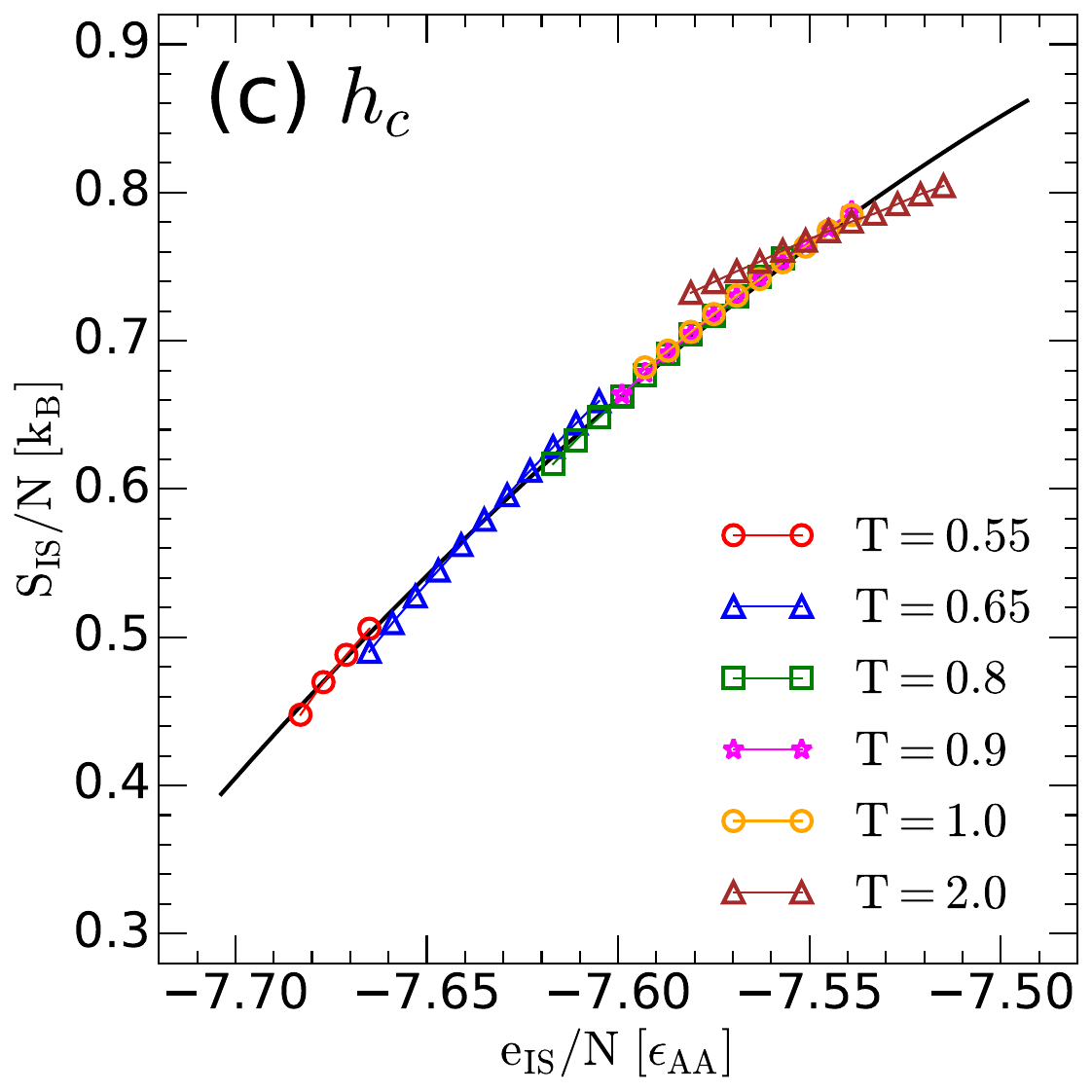}
	}    
	\caption{(a) Configurational entropy of the LJBMs as a function of the IS energy $e_{IS}$, $S_{IS}(e_{IS})$ [solid black line; taken from Fig.~5(a) of the main manuscript]. The symbols correspond to the expression on the right-hand-side of Eq.~\ref{SconfTEST2} for temperatures $T=0.435$ (red circles), $0.5$ (blue up-triangles), $0.6$ (green squares), $0.7$ (magenta stars) and $1.0$ (orange circles). Results at the different temperatures are obtained from ring-polymer molecular dynamics (RPMD) simulations with $h=h_a$ [datasets are shifted by a constant $c(T)$; see text]. (b) Same as (a) for $h=h_b$, and for temperatures $T=0.475$ (red circles), $0.55$ (blue up-triangles), $0.7$ (green squares), $0.8$ (magenta stars), $1.0$ (orange circles), and  $2.0$ (maroon up-triangles). (c) Same as (a) for $h=h_c$, and for temperatures $T=0.55$ (red circles), $0.65$ (blue up-triangles), $0.8$ (green squares), $0.9$ (magenta stars), $1.0$ (orange circles), and  $2.0$ (maroon up-triangles). In contrast to Fig. 6 of the main manuscript, the orange symbols in (a) and (b) [T=1.0] deviate slightly from the black line.  Similarly, the blue triangles in (c) [T=0.65] deviate slightly from the black line. }
	\label{fig:validate-Sconf-HA}
\end{figure}

That anharmonic contributions for the PEL formalism ($F_{vib}^{anh}(N,V,T)$) are relevant for the quantum LJBMs studied is evident from Figs.~2(a) and 2(b) of the main manuscript.  One may wonder, how relevant the anharmonic contributions are for the configurational entropy $S_{IS}(N,V,e_{IS})$ given in Eq.~10.   To address this point, we test whether Eq.~34 holds when anharmonicities are removed [$F_{vib}^{anh}(N,V,T) \rightarrow 0$].  Specifically, here we test whether Eq.~10 satisfies the following expression,
	\begin{eqnarray}
		S_{IS}(T,e_{IS})/k_B = \ln \left( \mathrm{P}(T,e_{IS}) \right) + 3 N n_b \ln(\beta \hbar \omega_0)  + {\cal S}(T,e_{IS}) + \beta e_{IS} + c(T)
		\label{SconfTEST2}
	\end{eqnarray}
As shown in Fig.\ref{fig:validate-Sconf-HA}, Eq.~\ref{SconfTEST2} is consistent with Eq.~10 for $h=0, h_a$.  However, a close comparison of Fig.~\ref{fig:validate-Sconf-HA} and Fig.~2(a)(b) of the main manuscript, shows that for $T=1.0$ (orange circles), Eq. 34 works slightly better than Eq.~\ref{SconfTEST2}.  As the quantum character of the LJBMs increases, small deviations are observed at lower temperatures.  While in Fig. 2(c) of the main manuscript the values of $S_{IS}$ at $T=0.65$  (blue triangles) obey Eq. 10 (black line), this is not the case when Eq.~\ref{SconfTEST2} is used [blue triangles and black line in Fig.~\ref{fig:validate-Sconf-HA}(c)].  Despite the slightly better performance of Eq. 34, compared to Eq.~\ref{SconfTEST2}, the anharmonic contributions to the $S_{IS}$ of the LJBMs studied seem to be small for $h=0, h_a, h_b, h_c$, but seem to become increasingly relevant as $h$ further increases ($h>h_c)$.

\clearpage






%